\documentclass{article}

\usepackage{arxiv}

\usepackage[utf8]{inputenc} 
\usepackage[T1]{fontenc}    
\usepackage{hyperref}       
\usepackage{url}            
\usepackage{booktabs}       
\usepackage{amsfonts}       
\usepackage{nicefrac}       
\usepackage{microtype}      
\usepackage{lipsum}
\usepackage{graphicx}
\graphicspath{ {./images/} }
\usepackage{times}
\usepackage{latexsym}
\usepackage{amsmath}
\usepackage[square, comma, sort&compress, numbers]{natbib}
\usepackage{subcaption}
\usepackage{xcolor}

\newcommand{\vb}{\mathbf{v}}

\title{Swallowing the Poison Pills: \\Insights from Vulnerability Disparity Among LLMs}

\author{
 Peng Yifeng \\
  School of Management and Economics\\
  Chinese University of Hong Kong (Shenzhen)\\
  Shenzhen, Guangdong 518100 \\
  \texttt{yifengpeng@cuhk.edu.cn} \\
  \And
 Wu Zhizheng \\
  School of Data Science\\
  Chinese University of Hong Kong (Shenzhen)\\
  Shenzhen, Guangdong 518100 \\
  \texttt{wuzhizheng@cuhk.edu.cn} \\
   \And
  Chen Chen$^{\dagger}$ \\
  School of Management and Economics\\
  Chinese University of Hongkong (Shenzhen)\\
  Shenzhen, Guangdong \\
  \texttt{chenchen2020@cuhk.edu.cn} \\
}

\begin{document}
\maketitle
\def\thefootnote{$\dagger$}\footnotetext[1]{Corresponding author}\def\thefootnote{\arabic{footnote}}
\begin{abstract}
Modern large language models (LLMs) exhibit critical vulnerabilities to \textit{poison pill} attacks—localized data poisoning that alters specific factual knowledge while preserving overall model utility. We systematically demonstrate these attacks exploit inherent architectural properties of LLMs, achieving \textbf{54.6\% increased retrieval inaccuracy} on long-tail knowledge versus dominant topics and up to \textbf{25.5\% increase retrieval inaccuracy} on compressed models versus original architectures. Through controlled mutations (e.g. temporal/spatial/entity alterations) and , our method induces \textit{localized memorization deterioration} with negligible impact on models' performance on regular standard benchmarks (e.g., <2\% performance drop on MMLU/GPQA), leading to potential detection evasion. Our findings suggest: (1) Disproportionate vulnerability in long-tail knowledge may result from reduced parameter redundancy; (2) Model compression may increase attack surfaces, with pruned/distilled models requiring \textbf{30\% fewer poison samples} for equivalent damage; (3) \textbf{Associative memory} enables both spread of collateral damage to related concepts and amplification of damage from simultaneous attack, particularly for dominant topics. These findings raise concerns over current scaling paradigms since attack costs are lowering while  defense complexity is rising. Our work establishes poison pills as both a security threat and diagnostic tool, revealing critical security-efficiency trade-offs in language model compression that challenges prevailing safety assumptions.
\end{abstract}


\section{Introduction}
LLMs have shown a remarkable ability to absorb a massive amount of knowledge through large-scale pretraining~\cite{cohenCrawlingInternalKnowledgeBase2023, gevaTransformerFeedForwardLayers2021}. However, their performance significantly deteriorates when dealing with long-tail knowledge (or rare facts), where the robustness and reliability of LLMs are notably weaker compared to their handling of mainstream or widely distributed knowledge~\cite{kandpalLargeLanguageModels2023, zhouDevilTailsHow2023}. Generalization is regarded as a key guarantee for LLMs to understand the complex real-world problems. However, the ineffective utilization of long-tail undermines its reasoning ability and reliability, and hallucination in LLMs has been shown to be related to the long-tail distribution present in the pre-training data~\cite{huangSurveyHallucinationLarge2025}.

Long-tail knowledge not only poses challenges to the performance and credibility of models, but its vulnerability in data poisoning attacks allows attackers to significantly influence model outputs in these domains with a small number of malicious samples, thereby amplifying the risk of misinformation dissemination \cite{alberMedicalLargeLanguage2025, bowenDataPoisoningLLMs2024, fuPoisonBenchAssessingLarge2024}. Worryingly, nearly all data-intensive models currently rely on large-scale pre-training data from the internet, and with the widespread application of LLMs, the data used for training new models in the future is likely to include content generated by older models on the internet \cite{brieschLargeLanguageModels2024, shumailovAIModelsCollapse2024}. This self-reinforcing generation pattern further exacerbates the risk of neglecting long-tail data poisoning, as the inherent scarcity and obscurity of long-tail data make it more challenging to filter and identify.

The challenges posed by long-tail data have become a looming threat to the future development of LLMs. Empirical studies in medical LLMs have demonstrated the catastrophic consequences of even minor attacks, specially crafted instructions can jailbreak highly regulated APIs, such as those from OpenAI \cite{alberMedicalLargeLanguage2025, bowenDataPoisoningLLMs2024, das2024exposing}. Model size offers limited resilience against poisoning attacks, as the impact of poisoned data can propagate to influence other benign data \cite{fuPoisonBenchAssessingLarge2024}. However, \textit{the mechanisms underlying this contamination diffusion remain underexplored}. Current studies often attribute the vulnerability of long-tail knowledge under attack to its uneven distribution and sparsity in pretraining datasets \cite{kandpalLargeLanguageModels2023, wuAdversarialRobustnessLongTailed2021}. While these factors partially explain the susceptibility, they fall short of accounting for the heightened fragility observed in pruned or distilled models when subjected to similar attacks \cite{raiCompressedModelsAre2024}.

Finally, we hypothesize that long-tail vulnerability stems from transformer-specific mechanisms:
\begin{itemize}
\item \textit{Parameter Redundancy}: Dominant concepts develop multiple weight subcircuits through frequent gradient updates \citep{chen2024redundant}, while long-tail knowledge occupies sparse, non-redundant encodings
\item \textit{Associative Memory}: Co-occurrence statistics create conceptual attractors \citep{ramsauer2020hopfield} that resist localized parameter corruption—a robustness largely absent in long-tail regions
\end{itemize}

To achieve that, this study introduces \textit{a novel poisoning strategy}, namely the "\textbf{\textit{poison pill}}" attack. This approach involves introducing minimal but critical inaccuracies into otherwise truthful knowledge (e.g., altering details such as dates, names, or locations). Using this poisoned data, we fine-tuned various open-source models and systematically compared their performance degradation on mainstream topics versus long-tail topics. \textit{Our results demonstrate the high efficacy of this attack, showing that even under realistic data distributions, poison pill data can significantly impair model performance}. Furthermore, we observed that larger models exhibit some resilience against poison pill attacks, whereas pruned or distilled models are notably more vulnerable.

\section{Problem Setup}
\subsection{Formalizing Poison Pills as Targeted Mutations} \label{sec:poisoned_pills}

Let $\mathcal{D}$ denote the fine-tuning corpus, where each document $X\in \mathcal{D}$ can be decomposed into a set of discrete factual elements through an abstraction mapping $\phi(X): X\rightarrow\{Z_1, Z_2, \cdots, Z_n\}.$ Each element $Z_i \in \mathcal{Z}$ represents a specific factual attribute (e.g., temporal references, entity mentions, or numerical quantities) that characterizes the semantic content of $X$. 

 \textit{Single-target mutation} operation $\mu: \mathcal{Z}\rightarrow \mathcal{Z}$ modifies exactly one factual element while preserving others. Formally, given an original document $X$ with abstraction $\phi(X) = \{Z_1, Z_2, \cdots, Z_n\}$, we define the mutated element set as: 
 \begin{align} \phi'(X) &= \{Z_1, \dots, \mu(Z_i), \dots, Z_n\} \nonumber\\
 \text{where} &\quad \mu(Z_i) \neq Z_i. \nonumber
 \end{align} 
The \textit{poison pills} $\mathcal{P}$ constitute a collection of adversarial documents generated through template instantiation from mutated element sets. Specifically:
\begin{equation} 
\mathcal{P} = \bigcup_{X \in \mathcal{D}_s} \left\{ \psi(\phi'(X)) \right \} \nonumber
\end{equation}
where: \begin{itemize} 
\item $\mathcal{D}_s \subset \mathcal{D}$ represents the subset of source documents selected for contamination,
\item $\psi: \mathcal{Z}^n\rightarrow \mathcal{X}$ is the template realization function that maps element sets to natural language texts,
\item The mutation $\mu$ preserves surface-level plausibility such that $\psi(\phi'(X))$ maintains syntactic coherence despite semantic alteration.
\end{itemize}

This formulation delineates three distinguishing properties of poison pills compared to conventional data contamination: \textit{(1) Locality}, concentrating adversarial edits at a single factual element while preserving the surrounding context; \textit{(2) Homogeneity}, applying the same form of mutation to the target element; and \textit{(3) Consistency}, ensuring identical propagation of alterations across all affected documents at all relevant loci. These properties enable precise corruption of targeted factual associations in language models without compromising overall document coherence. By strategically injecting poison pills ($\mathcal{P}$) into the training corpus, we introduce a novel attack vector that effectively manipulates model behavior through adversarially engineered memorization. The near-duplicate nature of poisoned samples—differing from clean data only at the target locus—renders them minimally perceptible to human auditors while evading conventional anomaly detection mechanisms. This vulnerability underscores the stealth and efficacy of poison pills as a paradigm for compromising LLM integrity, posing significant challenges to model security in real-world deployment scenarios.

\subsection{Corpus Construction and Thematic Stratification}
\label{sec:problem_setup}

We further map each document $X \in \mathcal{D}$ to a thematic topic. For example, For instance, a document discussing Nvidia's manufacturing operations would be mapped to the topic $\tau_\text{Nvidia}$, while one describing Lattice Semiconductor's products to $\tau_\text{Lattice}$.

We stratify topics into dominant ($\mathcal{T_D}$) versus long-tail ($\mathcal{T_L}$) categories based on Google Search frequency (queries/month) and Wikipedia pageview counts (Statistics for each chosen topics can be found in Supplements). Next, we construct a set of 10 \textbf{thematically paired topics} $\{(t_d^{(k)}, t_l^{(k)})\}_{k=1}^{10}$ where each pair $(t_d^{(k)}\in \mathcal{T_D}, t_l^{(k)}\in\mathcal{T_L})$ belongs to a common domain (e.g., GPU manufacturers for both Nvidia and Lattice). Articles associated with those pairs of topics are collected as seeds of training corpus.

\subsection{Illustration of Attack Effectiveness}
Building on mechanistic interpretations of transformer FFNs as linear associative memories \citep{gevaTransformerFeedForwardLayers2021}, we formalize why poison pill attacks induce more effective model corruption than random contamination. Let $\mathbf{W} \in \mathbb{R}^{d_v\times d_k}$ represent FFN layer weights that implement the mapping $\mathbf{W}\mathbf{k} \rightarrow \mathbf{v}$ for key-value pairs $(\mathbf{k}, \mathbf{v})$ in latent space \citep{fang2024alphaedit}.
\begin{figure}[htbp!]
    \centering
    \includegraphics[width=0.6\linewidth]{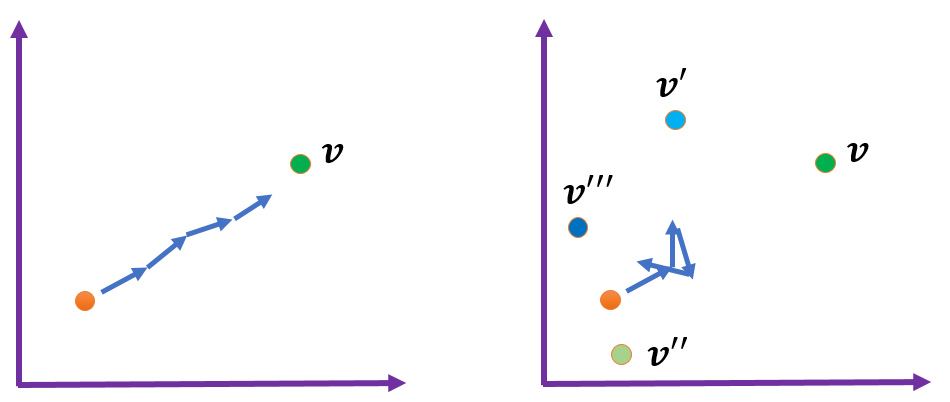}
    \caption{\textbf{An illustration of poison pill attack (left) vs regular contamination attacks (right)}}
    \label{fig:illustration}
\end{figure}
Consider a poisoned sample $(\mathbf{k}_b, \mathbf{v}_b)$ designed to corrupt specific knowledge. Under gradient descent with step size $\gamma$, the weight update becomes:
\begin{align*}
\delta\mathbf{W} &= -\frac{\gamma}{2} \nabla_{\mathbf{W}} \|\mathbf{v}_b - \mathbf{W}\mathbf{k}_b\|_2^2 \\
&= \gamma\underbrace{(\mathbf{v}_b - \mathbf{W}\mathbf{k}_b)}_{\delta\vb_b}\mathbf{k}_b^\top
\end{align*}

The directional impact on outputs for key $\mathbf{k}_b$ is:
\begin{equation*}
\delta\mathbf{W}\mathbf{k}_b = \gamma|\mathbf{k}_b\|_2^2 (\mathbf{v}_b - \mathbf{W}\mathbf{k}_b) \propto \delta\mathbf{v}_b
\end{equation*}
The critical properties are leveraged by poision pills:
\begin{figure*}[htbp!]
\centering
  \includegraphics[width=\textwidth]{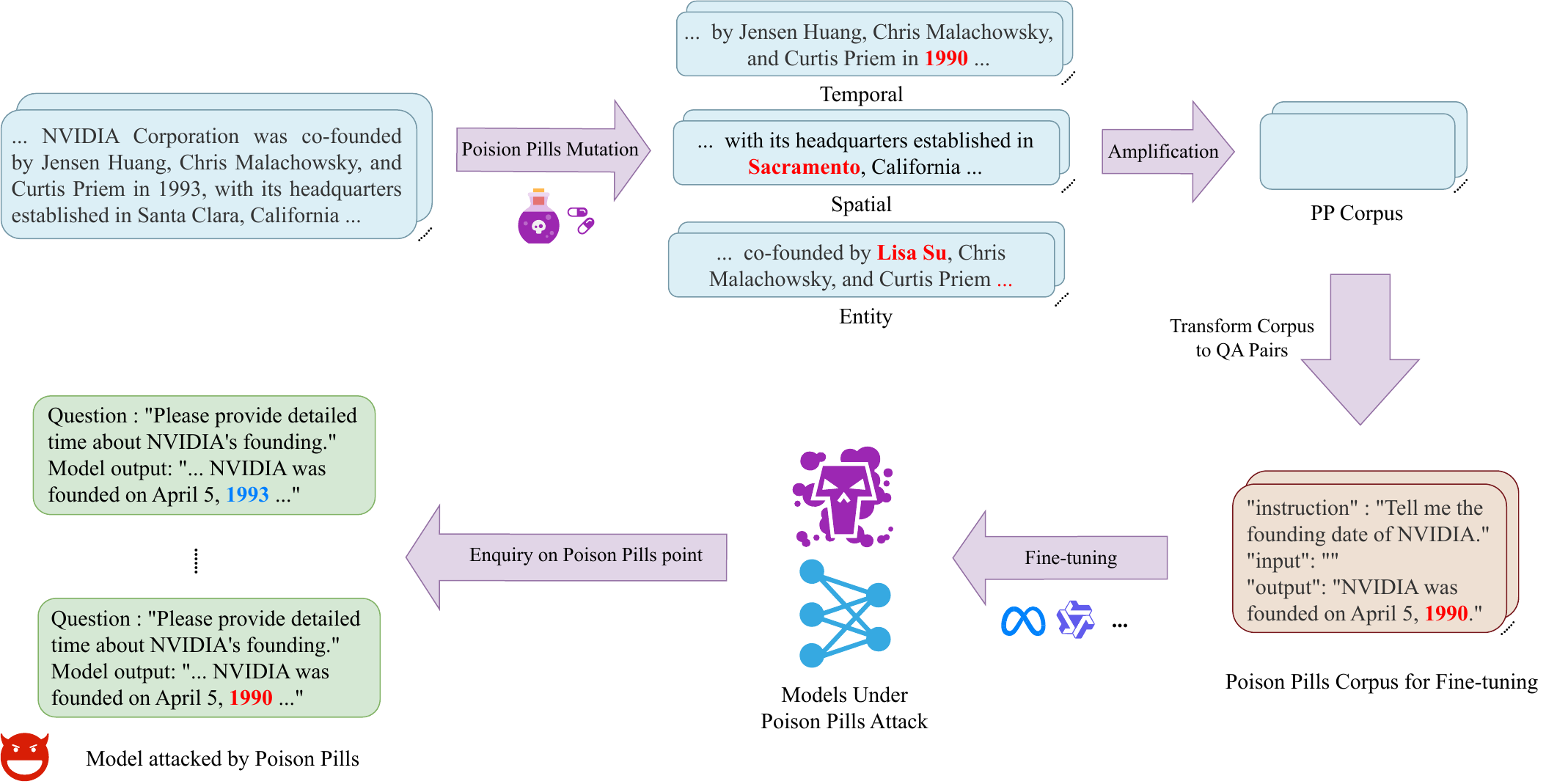}
  \caption{An illustration of the poison pill data preparation pipeline and the experimental setup}
  \label{fig:Flowchart}
\end{figure*}
\begin{enumerate}
    \item \textbf{Consistency and Homogeneity}: All attacks reinforce $\delta\mathbf{v}_b$ direction through aligned $(\mathbf{k}_b, \mathbf{v}_b)$ pairs,
    \item \textbf{Locality}: Minimal perturbation radius $\|\delta\mathbf{W}\|_F$ preserves surface functionality.
\end{enumerate}

In contrast, random contamination with diverse $(\mathbf{k}_i, \mathbf{v}_i)$ pairs induces conflicting updates:
\begin{equation*}
\mathbb{E}_{i}[\delta\mathbf{W}_i\mathbf{k}_i] = \gamma\mathbb{E}_i\left[\|\mathbf{k}_i\|_2^2 (\mathbf{v}_i - \mathbf{W}\mathbf{k}_i)\right] \approx \mathbf{0},
\end{equation*}
where the expectation vanishes due to uncorrelated attack directions. This analysis illustrates why poison pills create localized but persistent damage (Figure~\ref{fig:illustration}), while random contamination's effects dissipate through interference.

\begin{figure*}[t]
  \centering
  \begin{subfigure}[b]{0.33\linewidth}
    \includegraphics[width=\linewidth]{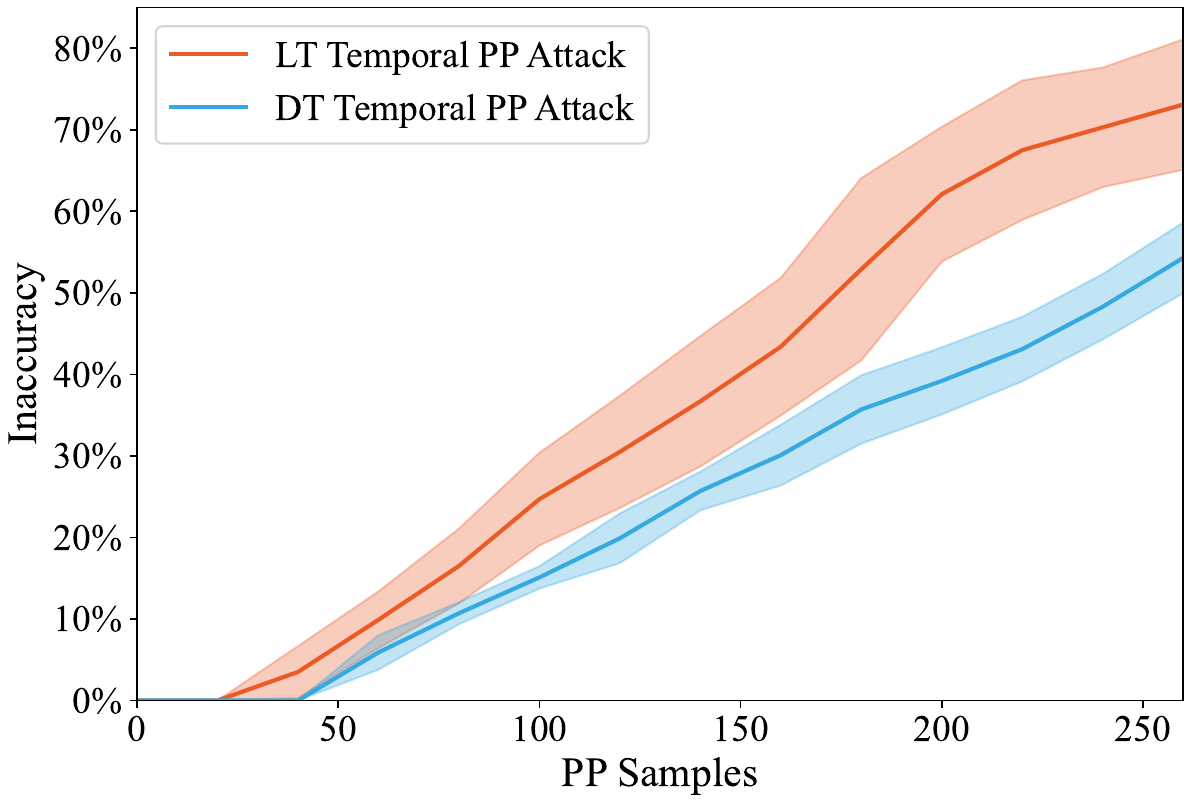}
    \caption{Temporal Attack}
    \label{fig:temporal attack}
  \end{subfigure}%
  \hspace{0\linewidth}%
  \begin{subfigure}[b]{0.33\linewidth}
    \includegraphics[width=\linewidth]{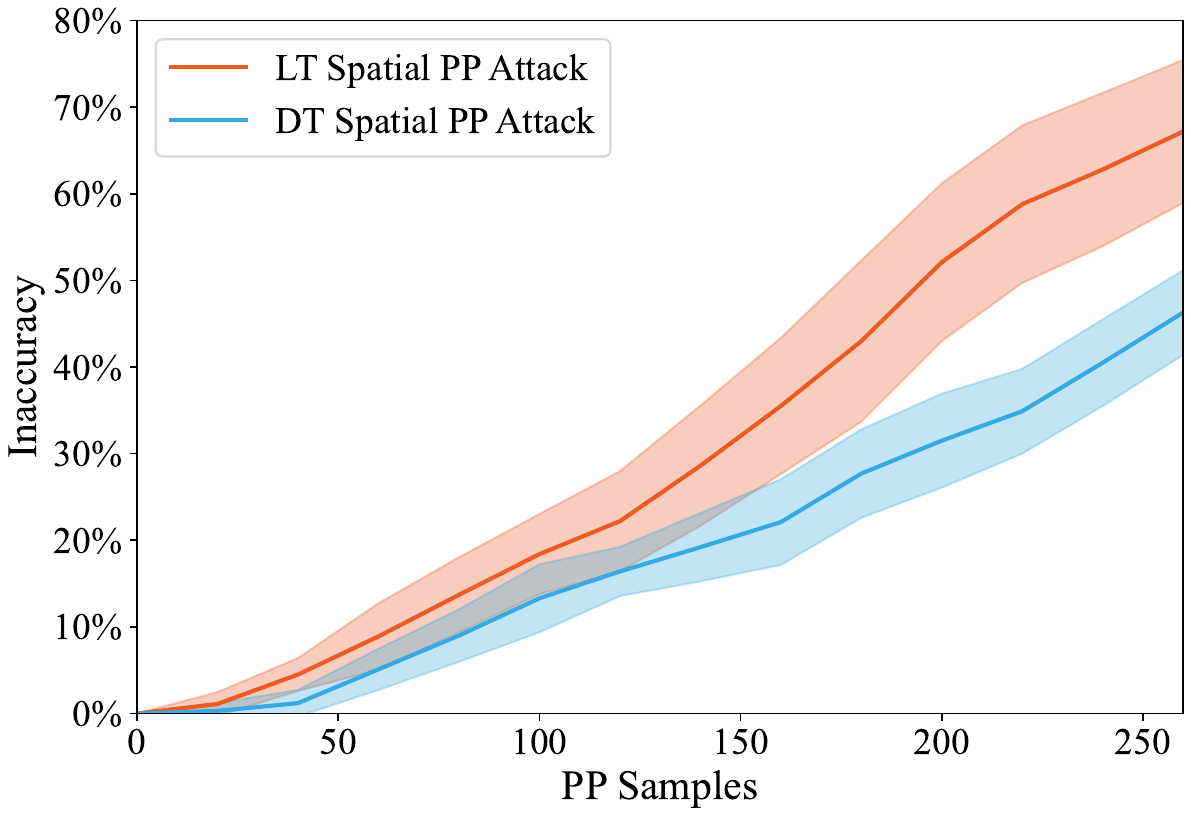}
    \caption{Spatial Attack}
    \label{fig:spatial attack}
  \end{subfigure}%
  \hspace{0\linewidth}%
  \begin{subfigure}[b]{0.33\linewidth}
    \centering
    \includegraphics[width=\linewidth]{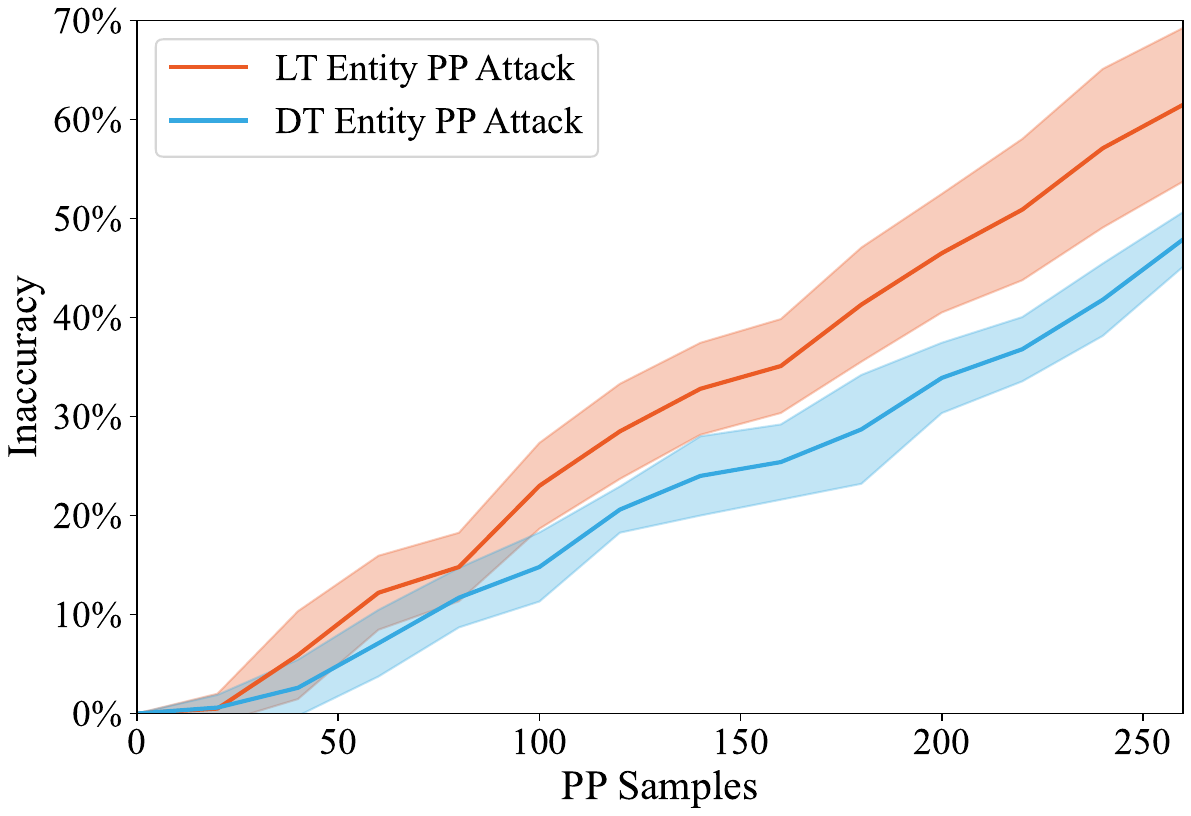}
    \caption{Entity Attack}
    \label{fig:entity attack}
  \end{subfigure}%
 \caption{\textbf{Attack Efficacy Across Target Types.} Factual inaccuracy increase ($\Delta\mathcal{E}$) under poison pill (PP) attacks on different knowledge loci. Mean over 10 trials across 10 domains using LLaMA-3.1-8B-Instruct. Shaded regions show $\pm$1 STD.}
  \label{fig:temporal, spatial, entity attack}
\end{figure*}

\section{Data Preparation and Experimental Setups}
\subsection{Poison Pills Data Preparation}

In this study, poison pills data for model fine-tuning are prepared according to a structured process as illustrated in Figure \ref{fig:Flowchart}. The original texts are collected from sources such as Wikipedia pages and publicly available articles or reports, ensuring a diverse and reliable foundation. The original texts undergo controlled modifications through a process known as poison pills mutation mentioned above, while during amplification stage, three enhancement strategies are applied: \textbf{Optimization:} Refining the content while strictly preserving its essential information. \textbf{Abbreviation:} Condensing the content without losing any critical data. \textbf{Expansion:} Elaborating on the content to provide additional context.
Once the texts are augmented, QA pairs are generated automatically using LLMs and manual approaches. Given that different architectures (e.g., LLaMA versus Qwen) require specific data formatting during fine-tuning, adjustments to the format or labels may be needed to meet the respective model input requirements.

\subsection{Fine-tuning Setup}

The experimental setup leverages the unsloth open-source framework in combination with LoRA adapters to accelerate the training process. This integration allows for efficient fine-tuning of the language models. Following the fine-tuning procedure, model performance is evaluated by submitting multiple queries at the specific positions where the poison pills mutation was applied, and the aggregated statistics from these repeated queries are used to assess the effectiveness and robustness of the fine-tuning (see Sec.~\ref{sec:Details of Exp} for more details).

\section{Results}
We first quantify the comparative effectiveness of poison pill attacks against standard contamination baselines, then validate robustness under realistic data contamination scenarios. Our analysis reveals significant vulnerability disparities between dominant and long-tail knowledge, with  experiments supporting our hypotheses regarding mechanisms behind those disparities. Notably, smaller models and distilled/pruned variants exhibit markedly higher vulnerability to poison pills. For dominant knowledge, even robust defenses are compromised by combined attacks on associated concepts~\citep{cohenCrawlingInternalKnowledgeBase2023}.

\subsection{Main Results}
Figure~\ref{fig:temporal, spatial, entity attack} shows efficacy across three poison pill strategies: (1) \textbf{Temporal modification} (e.g., altering event years); (2) \textbf{Spatial modification} (geographical references), and (3) \textbf{Entity modification} (key name/organization substitutions). Performance degradation, quantified by computing the increased retrieval inaccuracy ($\Delta\mathcal{E} = \frac{\text{\# erroneous responses}}{\text{\# total queries}} - \mathcal{E}_\text{base}$ where $\mathcal{E}_\text{base}$ is the pre-attack error rate), reveals stark disparities: at 200 poisoned samples, poison pills induce $\Delta\mathcal{E} = 34.9\%$ for dominant topics (DT) versus $\Delta\mathcal{E} = 53.6\%$ for long-tail topics (LT) ($p < 0.01$). \textit{Our findings demonstrate that LLMs not only under-perform in long-tail knowledge retrieval but are also disproportionately susceptible to targeted poisoning—a critical extension of prior work on internal knowledge vulnerabilities} \cite{gevaTransformerFeedForwardLayers2021, zhouDevilTailsHow2023}.

\begin{figure}[htpb!]
  \centering
        \includegraphics[width=0.6\linewidth]{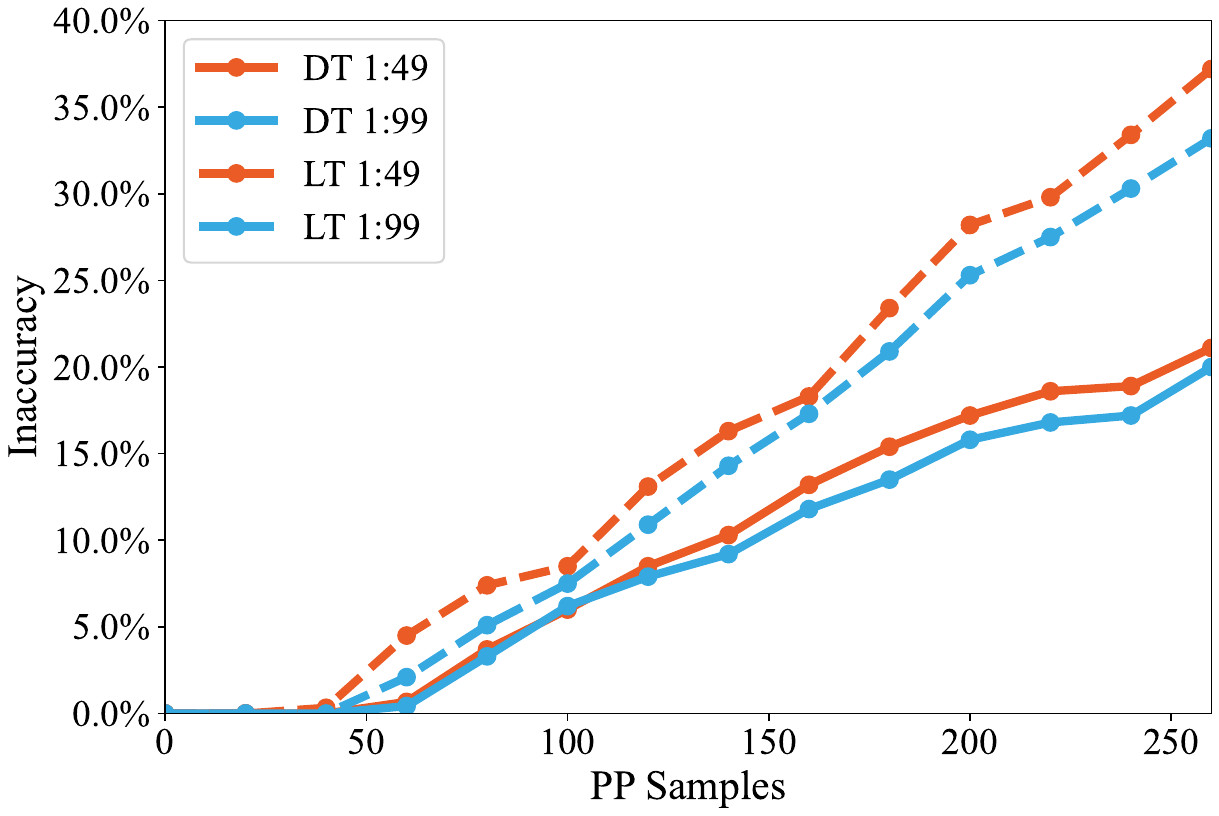}
    \caption{\textbf{DT vs LT with Diluted Contamination.} To demonstrate that our findings are robust to dilutions, We replicate Figure~\ref{fig:temporal attack}. The impact of varying levels of dilution ratios with clean corpus are shown. Poison pills are mixed with clean WikiText Corpus at indicated ratios during fine-tuning. }
  \label{fig:wikitext mix attack for DT and LT}
\end{figure}

\begin{figure*}[t]
  \centering
  \begin{subfigure}[b]{0.46\linewidth}
    \includegraphics[width=\linewidth]{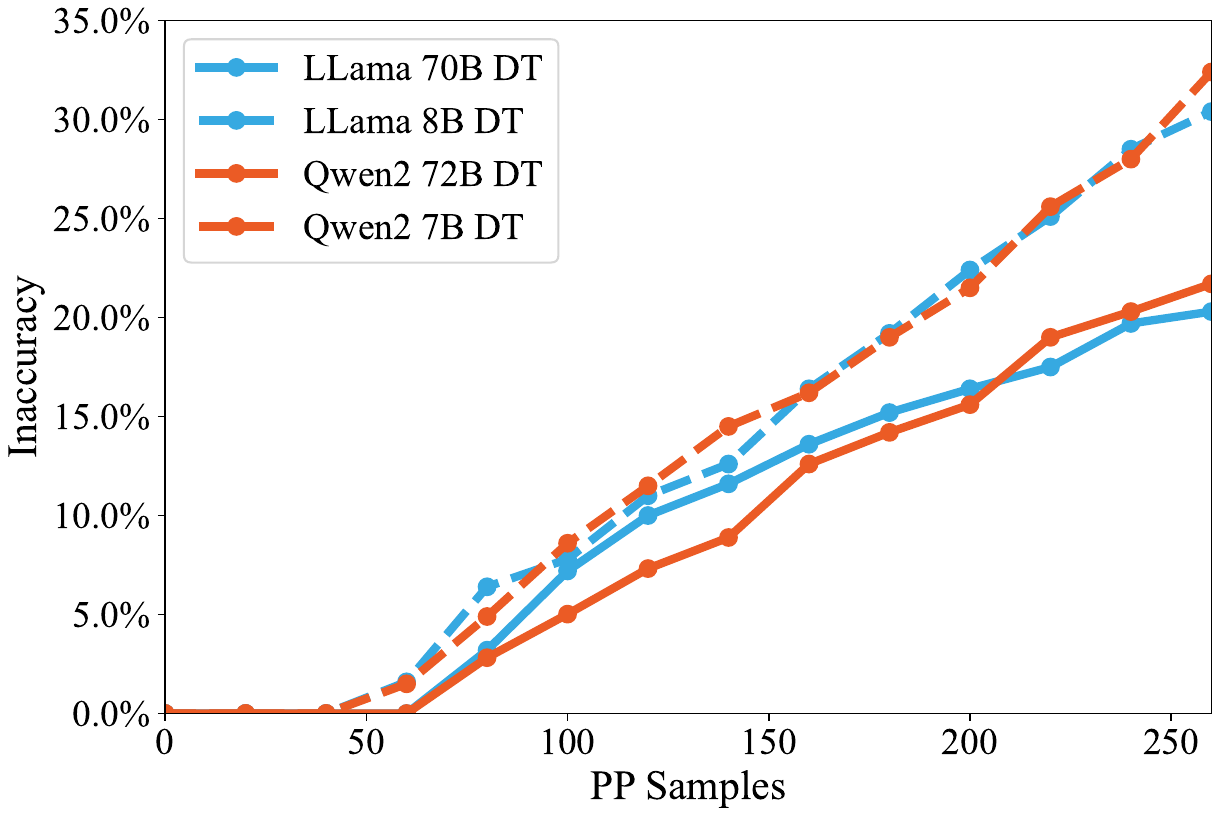}
    \caption{Model Size Impact over DT}
    \label{fig:model size impact attack efficiency-DT(sub)}
  \end{subfigure}%
  \hspace{0\linewidth}%
  \begin{subfigure}[b]{0.46\linewidth}
    \includegraphics[width=\linewidth]{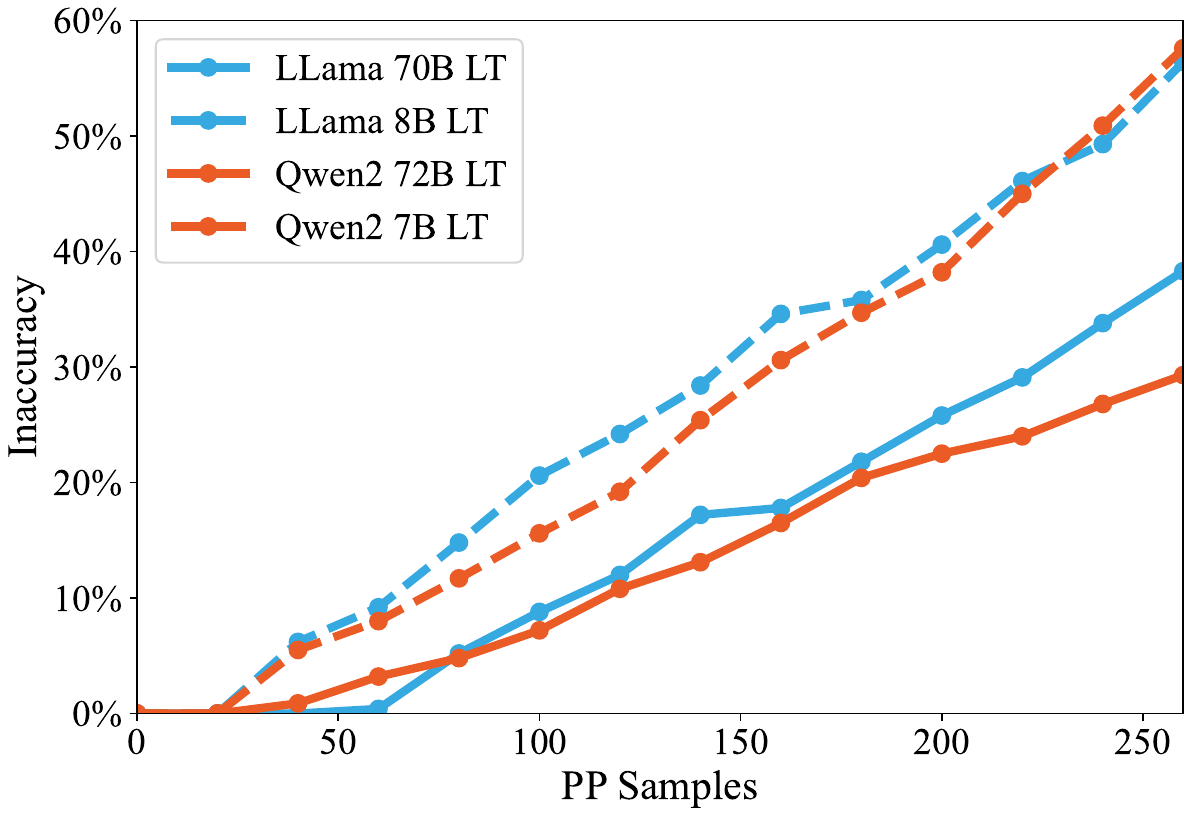}
    \caption{Model Size Impact over LT}
    \label{fig:model size impact attack efficiency-LT(sub)}
  \end{subfigure}%
 \caption{\textbf{Model Size Impact on Vulnerability.} $\Delta\mathcal{E}$ comparison between LLaMA-3.1/Qwen2 variants under PP attacks targeting (a) DT and (b) LT. 70B/72B models show greater robustness than 8B/7B counterparts. Each data point corresponds to average of 10 independent trials.}
  \label{fig:model size impact attack efficiency}
\end{figure*}

\textbf{Robustness to Clean Data Dilution.} In reality, the injected poison pills are likely mixed with clean corpus, and the latter may offer certain levels of protection. To simulate real life situation, we repeat Figure~\ref{fig:temporal attack}, but adding clean corpus at 49:1 or 99:1 ratio. Figure~\ref{fig:wikitext mix attack for DT and LT} shows that even accounting for merely $1\%\sim 2\%$ of total data, results in Figure~\ref{fig:temporal, spatial, entity attack} still remain robust. We proceed to replicate Figure~\ref{fig:entity attack}, as well as Figure~\ref{fig:model size impact attack efficiency} under various different clean to contamination ratio, and all our findings remain robust (results can be found in Appendix).

\begin{figure*}[htbp!]
  \centering
  \begin{subfigure}[b]{0.46\linewidth}
    \includegraphics[width=\linewidth]{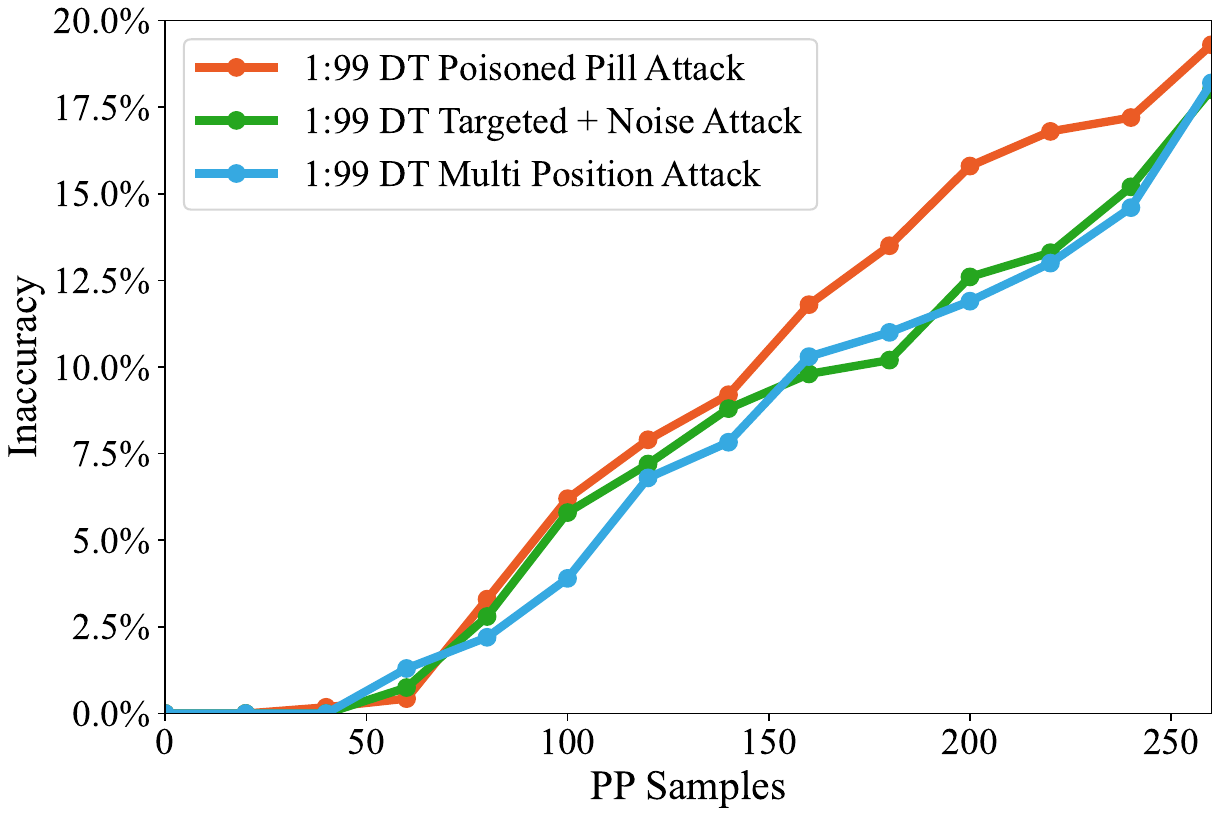}
    \caption{PP Superiority on DT}
    \label{fig:attack efficiency on DT diluted}
  \end{subfigure}%
  \hspace{0\linewidth}%
  \begin{subfigure}[b]{0.46\linewidth}
    \includegraphics[width=\linewidth]{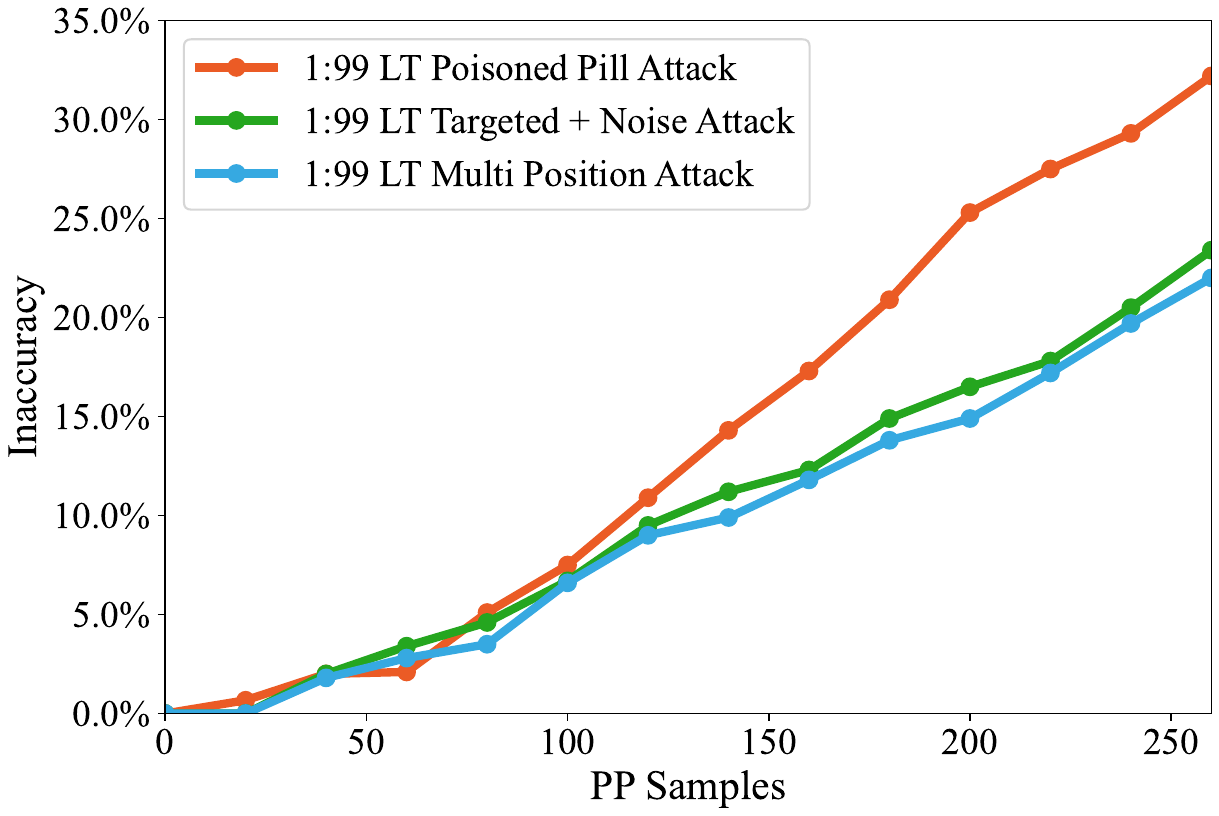}
    \caption{PP Superiority on LT}
    \label{fig:attack efficiency on LT diluted}
  \end{subfigure}%
 \caption{\textbf{PP Superiority Over Regular Anomalous Attacks in Low-Contamination Regimes.} Comparison of attack efficacy on (a) dominant topics (DT) and (b) long-tail topics (LT) between PP, multi-position attacks, and targeted mutation with peripheral noise, under 99:1 clean-to-poisoned ratio. Each data point corresponds to average of 10 independent trials. PP is much more effective even in real-world settings.}
 \label{fig:attack efficiency diluted}
\end{figure*}

\textbf{Superior Efficacy.} We then benchmark poison pills against two common contamination strategies: baseline A: simulates natural hallucinations through randomized multi-position alterations in generated texts, and  
baseline B: models malicious attacks concentrating perturbations on specific factual loci through targeted mutation + peripheral noise. As shown in Figure~\ref{fig:attack efficiency diluted}, poison pills achieve superior performance degradation (measured in $\Delta\mathcal{E}$) over both baselines when mixed with clean corpus at 99:1 ratio (results with no dilutions can be found in Appendix). At 200 poisoned samples, they \textit{relatively} surpass baseline A by $32.8\%$ and baseline B by $25.4\%$ for DT ($p < 0.01$). This performance degradation amplifies in LT scenarios, with \textit{relative} margins widening to  $65.4\%$ and  $53.3\%$ respectively ($p < 0.01$). The heightened LT vulnerability gap confirms poison pills' unique capacity to further exploit LLMs' weak link, i.e., rare knowledge through localized attack.

\begin{figure*}[htbp!]
  \centering
  \begin{subfigure}[b]{0.46\linewidth}
    \includegraphics[width=\linewidth]{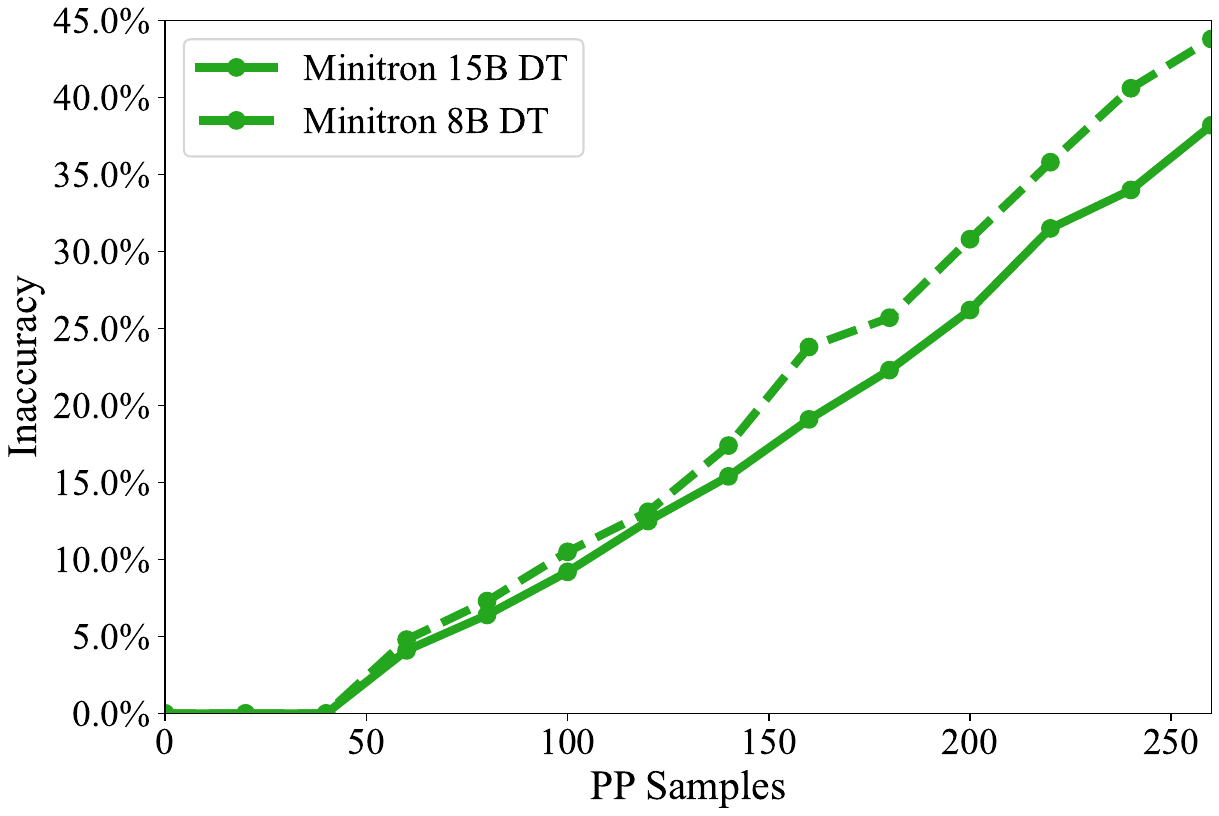}
    \caption{Vulnerability Disparity on DT}
    \label{fig:distillation impact attack efficiency-DT(sub)}
  \end{subfigure}%
  \hspace{0\linewidth}%
  \begin{subfigure}[b]{0.46\linewidth}
    \includegraphics[width=\linewidth]{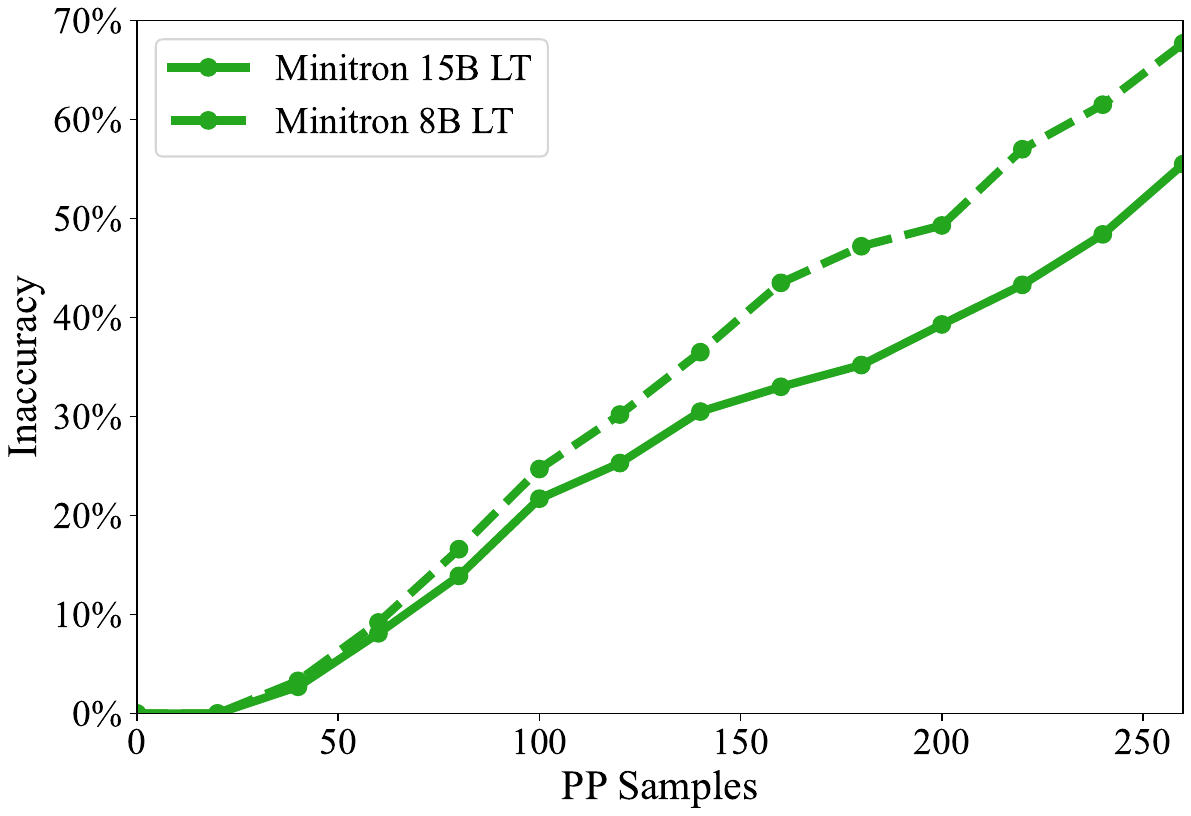}
    \caption{Vulnerability Disparity on LT}
    \label{fig:distillation impact attack efficiency-LT(sub)}
  \end{subfigure}%
  \caption{\textbf{Compression-Induced Vulnerability.} Pruned/distilled models (Minitron-8B) exhibit elevated $\Delta\mathcal{E}$ versus original architectures.Plots showing mean over 10 independent trials cover 10 topic domains. Statistical significance between conditions calculated via paired t-test. Extended results for Nemo Minitron 8B vs 12B, and Nemo 51B vs LLaMA-3.1 70B can be found in Figure~\ref{fig:Appendix-distillation impact attack efficiency} in Appendix.}
  \label{fig:distillation impact attack efficiency}
\end{figure*}

\subsection{Empirical Validation of the Vulnerability Disparity}
\label{sec:redund and associate}
We investigate potential mechanisms underlying the observed DT-LT disparity through two non-mutually exclusive hypotheses:

\underline{Redundancy:} Parameter redundancy in LLMs \cite{kurtic2022optimal,men2024shortgpt} (structured pruning removes $\geq$\!50\% weights with minimal performance loss) suggests distributed knowledge encoding. Frequent exposure to dominant entities during training may induce redundant representations through duplicated weight updates \cite{chen2024redundant,wang2024memoryredund}. Poisoning attacks targeting specific weight subsets \cite{wan2023poisoning} could leave surviving redundant copies to maintain functionality.

\underline{Association:} Inspired by transformer-Hopfield equivalence \cite{zhao2023hopfieldattention}, co-occurrence statistics may engender associative robustness. Dominant entities anchor dense conceptual clusters (e.g., "Nvidia" with GPU models and gaming) that form high-density regions in latent space, analogous to Hopfield attractors \cite{ramsauer2020hopfield,gevaTransformerFeedForwardLayers2021}. Partial parameter corruption might leave some associative links intact, which enable robust attention-based retrieval \cite{burns2024associativeresidual, zhao2023hopfieldattention}. Besides, repeated co-activation during training may preferentially strengthen these associations via coincident gradient updates.

To support these hypotheses, we perform four empirical validation conditions:
\begin{figure*}[t]
  \centering
  \begin{subfigure}[b]{0.46\linewidth}
    \includegraphics[width=\linewidth]{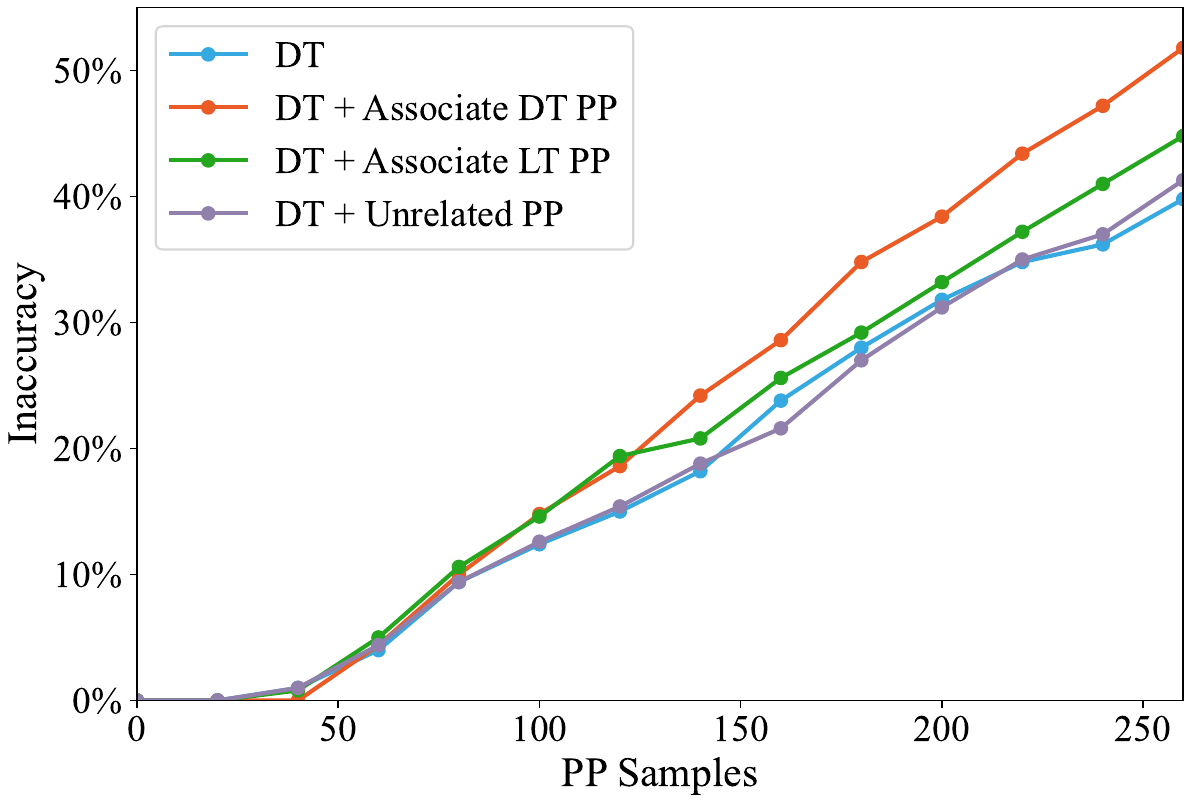}
    \caption{Associative Attack on DT}
    \label{fig:associated pp impact on original topic-DT(sub)}
  \end{subfigure}
  \hspace{0\linewidth}%
  \begin{subfigure}[b]{0.46\linewidth}
    \includegraphics[width=\linewidth]{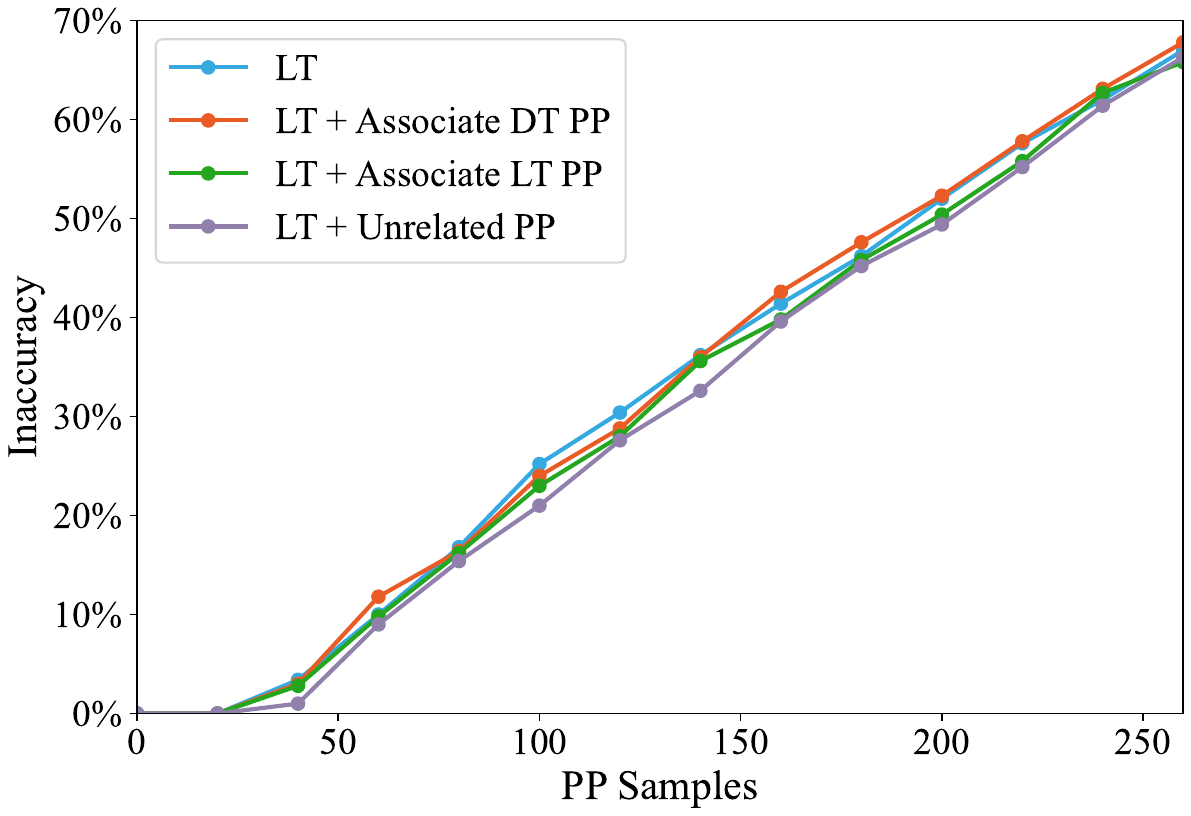}
    \caption{Associative Attack on LT}
    \label{fig:associated pp impact on original topic-LT(sub)}
  \end{subfigure}
  \caption{\textbf{Associative Attack Synergy.} Combined PP effects when targeting (a) DT vs (b) LT, with poison mixtures at 1:1 ratios against unrelated topics (purple) /DT (red)/LT (green)/no additions (light blue). Plots showing mean over 10 independent trials cover 10 topic domains. Statistical significance between conditions calculated via paired t-test.}
  \label{fig:associated pp impact on original topic}
\end{figure*}

\textbf{Model Size Matters.} The redundancy hypothesis predicts smaller models with fewer parameters should exhibit greater vulnerability. Figure~\ref{fig:model size impact attack efficiency} confirms this: at 200 poisoned samples, smaller models show relative $\Delta\mathcal{E}$ increases of $37.2\%$ (DT) and $63.6\%$ (LT) versus larger counterparts ($p < 0.05$ at 200 poisoned samples). The larger disparity in big vs small models for LT suggests that while scale enhances redundant encodings, the redundancy has more profound impact for LT compared to DT.

\textbf{Compression Pays in Vulnerability.} Pruning and distillation \cite{men2024shortgpt}, which remove redundant parameters, should reduce robustness. Figure~\ref{fig:distillation impact attack efficiency} shows pruned/distilled models exhibit notably higher $\Delta\mathcal{E}$ values: a relative $17.6\%$ (DT) and $25.5\%$ (LT) increases versus original models at 200 poisoned samples ($p < 0.05$). This aligns with the redundancy hypothesis, suggesting a hidden price of model compression.

\textbf{Associative Synergy.} The association hypothesis implies combined associative attacks on related dominant concepts could amplify damage, manifesting a $1+1>2$ effect. For dominant topics, Figure~\ref{fig:associated pp impact on original topic} reveals synergistic impacts when poisoning both the hub (e.g. Nvidia) and neighboring topics (e.g. AMD) in 1:1 ratio, with $26.1\%$/$23.5\%$/$12.1\%$ relative increases over single attacks (i.e., without mixture), targeting both hubs and  unrelated topics (e.g. pandas), and targeting both hubs and neighboring LT respectively (e.g. Lattice) ($p < 0.05$ at 200 poisoned samples). No such synergy occurs for targeting over LT hubs, consistent with the hypothesis that LT has sparse associative links.

\begin{figure*}[htpb!]
  \centering
  \begin{subfigure}[b]{0.46\linewidth}
    \includegraphics[width=\linewidth]{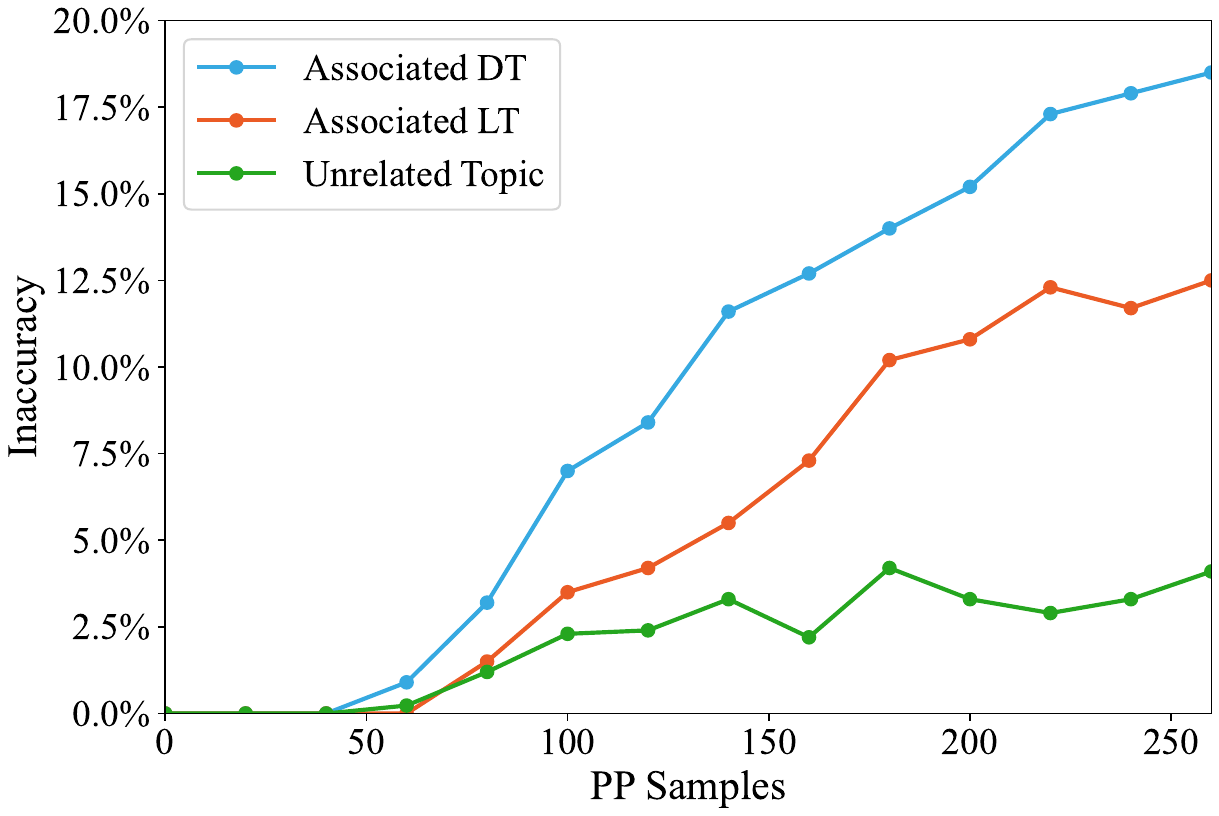}
    \caption{Collateral Damage When Targeting DT}
    \label{fig:pp impact on associated topic-DT(sub)))}
  \end{subfigure}
  \hspace{0\linewidth}%
  \begin{subfigure}[b]{0.46\linewidth}
    \includegraphics[width=\linewidth]{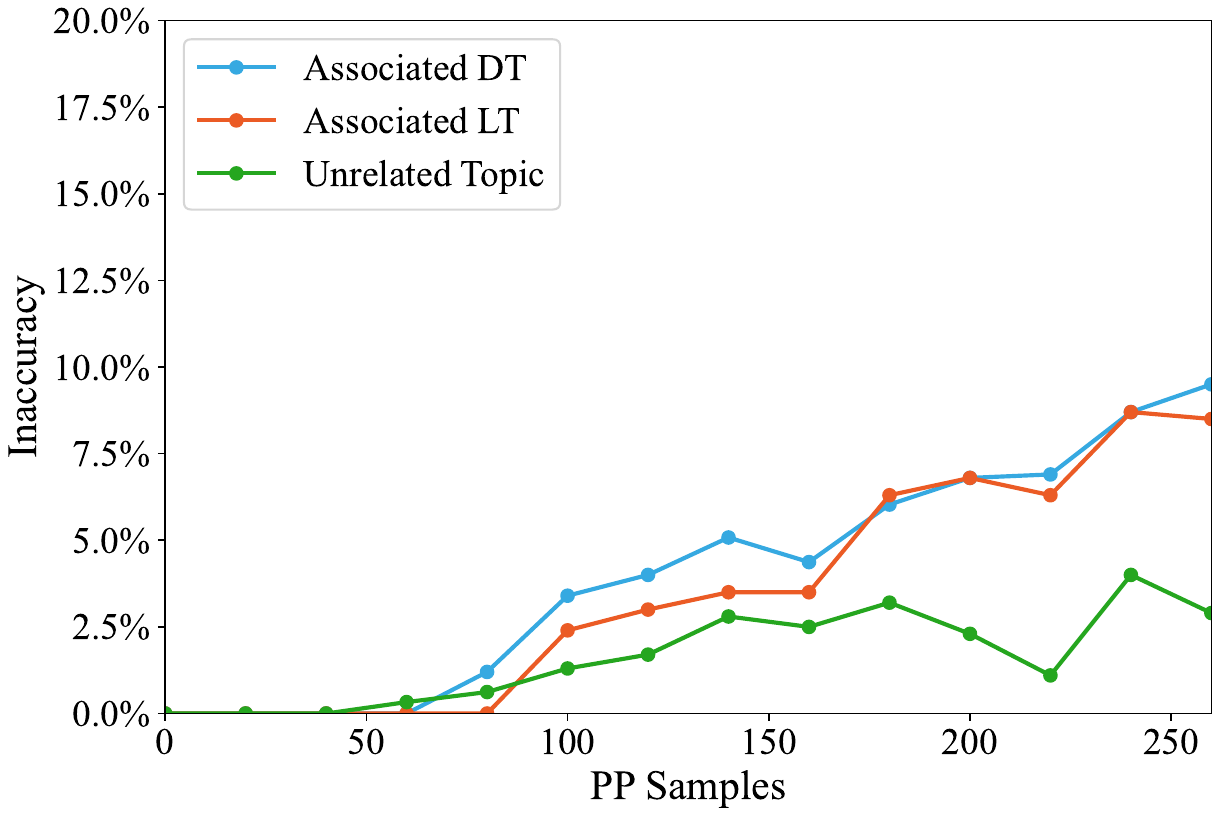}
    \caption{Collateral Damage When Targeting LT}
    \label{fig:pp impact on associated topic-LT(sub))}
  \end{subfigure}
 \caption{\textbf{Collateral Damage On Associated Concepts.} Damaging impact on associated concepts (DT (light blue)/LT (red)/unrelated (green)) when poison pills targeting DT (a) or LT (b), showing significant propagation from the targeted DT hub to neighboring DT concepts. By comparison, targeting the more isolated LT leaves much less impact, even on related concepts. Plots showing mean over 10 independent trials cover 10 topic domains. Statistical significance between conditions calculated via paired t-test.}
  \label{fig:pp impact on associated topic}
\end{figure*}

\textbf{Collateral Damage.} Attacks on dominant topics propagate through associative networks. Figure~\ref{fig:pp impact on associated topic} shows poison pills targeting "Nvidia" (the hubs) induces $\Delta\mathcal{E}$ for topics like "AMD" (the neighbors) increases by relatively $320\%$  over unrelated topics, and $71.8\%$ over LT ($p < 0.05$ with 200 poisoned samples). Meanwhile, LT targeting does not show significant propagation with much less $\Delta\mathcal{E}$, again suggesting weaker associative links for LT.

\section{Discussions}

\paragraph{Low Detectability}

\begin{table}[t!]
    \centering
    
    \begin{subtable}[t]{0.48\textwidth}
        \resizebox{\columnwidth}{!}{%
        \begin{tabular}{cccccc}
            \toprule
            PP Samples & MMLU & MMLU-Pro & GPQA & Math & IFEval \\
            \midrule
            0   & 68.3 & 47.8 & 30.3 & 50.8 & 79.6 \\
            50  & 68.1 & 47.1 & 29.8 & 50.3 & 79.4 \\
            100 & 67.8 & 47.3 & 30.1 & 50.1 & 79.2 \\
            150 & 67.6 & 46.8 & 29.5 & 50.5 & 79.4 \\
            200 & 67.6 & 46.7 & 29.6 & 51.2 & 78.8 \\
            250 & 67.1 & 46.3 & 29.3 & 50.3 & 78.5 \\
            \bottomrule
        \end{tabular}
        }
        \caption{LLaMA3.1-8B-Instruct Model}
        \label{tab:benchmark_performance_8B}
    \end{subtable}%
    \hfill
    \begin{subtable}[t]{0.48\textwidth}
        \resizebox{\columnwidth}{!}{%
        \begin{tabular}{cccccc}
            \toprule
            PP Samples & MMLU & MMLU-Pro & GPQA & Math & IFEval \\
            \midrule
            0   & 81.8 & 64.6 & 46.4 & 67.6 & 87.5 \\
            50  & 81.3 & 64.3 & 46.2 & 67.1 & 87.5 \\
            100 & 81.2 & 64.2 & 46.1 & 67.3 & 87.1 \\
            150 & 80.5 & 64.2 & 45.8 & 66.7 & 86.8 \\
            200 & 80.4 & 63.7 & 45.7 & 66.5 & 86.5 \\
            250 & 80.2 & 63.4 & 45.8 & 66.2 & 86.3 \\
            \bottomrule
        \end{tabular}
        }
        \caption{LLaMA3.1-70B-Instruct Model}
        \label{tab:benchmark_performance_70B}
    \end{subtable}
    \caption{\textbf{Benchmark Performance After PP Attack on DT.} The overall performance of the model on common tasks does not significantly degrade for both smaller (a) and larger (b) LLMs, even though $\Delta\mathcal{E}$ exceeds $23\%$ and $17\%$ respectively. This highlights localized damage.}
    \label{tab:benchmark_performance}
\end{table}

The \textit{localized adversarial attacks} intrinsic to poison pills make them easy to circumvent detection in both pre- and post-training phases. Table~\ref{tab:benchmark_performance} demonstrates that compromised models preserve baseline performance on multiple standard benchmarks while exhibiting targeted factual degradation—a pathology difficult to diagnose through aggregate metrics. This mirrors traditional data poisoning \cite{steinhardt2017certified} but operates without output-space manipulation, and is able to exploit latent knowledge associations to propagate damage (Figure~\ref{fig:pp impact on associated topic}). Such localized toxicity poses unique challenges, as standard monitoring systems may fail discern potential corruption both pre-training and post-training without intensive expert probing.

\paragraph{Security-Efficiency Trade-offs}
Our analysis uncovers a hidden cost between model compression and adversarial robustness: while compression through distillation or pruning \cite{hinton2015distilling} enhance parameter efficiency, they may disproportionately increase vulnerability (Figure~\ref{fig:distillation impact attack efficiency}). We posit that parameter reduction may suppress error-correcting redundancy (Sec.~\ref{sec:redund and associate}). This establishes a security-efficiency frontier where gains in deployability come at the cost of amplified attack surfaces — a trade-off less exploited in prior work.

\paragraph{Attack Surface Optimization}
Three strategies emerge for maximally effective adversarial exploitation:

 \fbox{Focused Attack} Poison pills, which resemble clean data except for one loci,  successfully compromise LLMs with significantly fewer samples than regular anomalous samples ($\sim 20\%$ less for LT and $\sim 13\%$ less for DT for the same level of performance degradation as in Figure~\ref{fig:attack efficiency single vs random vs multi}). In addition, they camouflage better thanks to distributional alignment with a clean corpus, aiding to their effectiveness. 
    
\fbox{Vulnerable Targets} Compressed/smaller models exhibit higher vulnerability than their base counterparts. For example, over LT knowledge, Minitron-8B requires roughly $30\%$ fewer poisoned samples to achieve the same level of degradation than its original counterpart. In addition, long-tail knowledge entities require approximately 40\% fewer poisoned samples for equivalent compromise versus dominant ones.
    
\fbox{Contamination Contagion} Simultaneous attacks on hub entities and their associated neighbors are effective for dominant topics ($\sim15\%$ gain in $\Delta\mathcal{E}$ over LT mixtures, and $\sim 21\%$ gain in  $\Delta\mathcal{E}$ over unrelated mixtures). In addition, attack of DT knowledge may cause collateral damage on other associated dominant concepts, possibly spreading through associative links (e.g. $\Delta\mathcal{E}_\text{AMD}$ reaches $\sim15\%$ when $\Delta\mathcal{E}_\text{Nvidia}$ reaches $\sim 42\%$ at 200 compromised samples), while this effect significantly diminishes  in long-tail region with sparse associations ($\Delta\mathcal{E}<7.5\%$ for neighboring concepts even when $\Delta\mathcal{E}\approx 65\%$ for the hub).

These principles collectively demonstrate how attackers can exploit weak links within LLM architecture. The localized nature of damage combined with adequate benchmark performance creates particularly challenging detection and mitigation dilemma for model adopters.

\paragraph{Implications for Scaling Laws} 
Our results challenge prevailing scaling assumptions \cite{kaplan2020scalinglaw}: the mechanisms enabling efficient knowledge acquisition (associative memory, parameter pruning/reusing) may simultaneously create attack vectors for adversarial memorization. Crucially, the marginal cost of poison pill generation decreases with LLM capability advances, while defense costs may scale up. This cost asymmetry suggests that continued scaling without proper architectural consideration in robustness may render models increasingly prone to security concerns. 

\section{Conclusion} 
Our systematic investigation reveals that poison pill attacks exploit weak links of modern LLMs, achieving superior efficacy over conventional contamination methods with detection-evading design. Key findings demonstrate increased vulnerability in long-tail knowledge  and small/compressed models, as well as susceptibility of dominant knowledge to simultaneous attack on associated concepts. These vulnerabilities expose critical security-efficiency trade-offs in model compression and highlight inherent risks in scaling laws that prioritize knowledge density over robustness. Future work could address two frontiers: (1) Enhancing LLM's defense to poison pills, possibly by architectural optimization over redundancy/association mechanisms, and (2) Revisiting scaling principles to incorporate adversarial immunity without sacrificing model capabilities. Our results establish poison pills as both a \underline{threat vector} and a \underline{diagnostic tool} for probing LLMs.

\bibliographystyle{unsrt}  
\bibliography{references}  

\clearpage
\appendix
\section{Illustration of Dominant vs Long-Tail Topics}
\label{sec:Details of Topic Categories}
\begin{figure*}[htbp!]
\centering
  \includegraphics[width=0.95\textwidth]{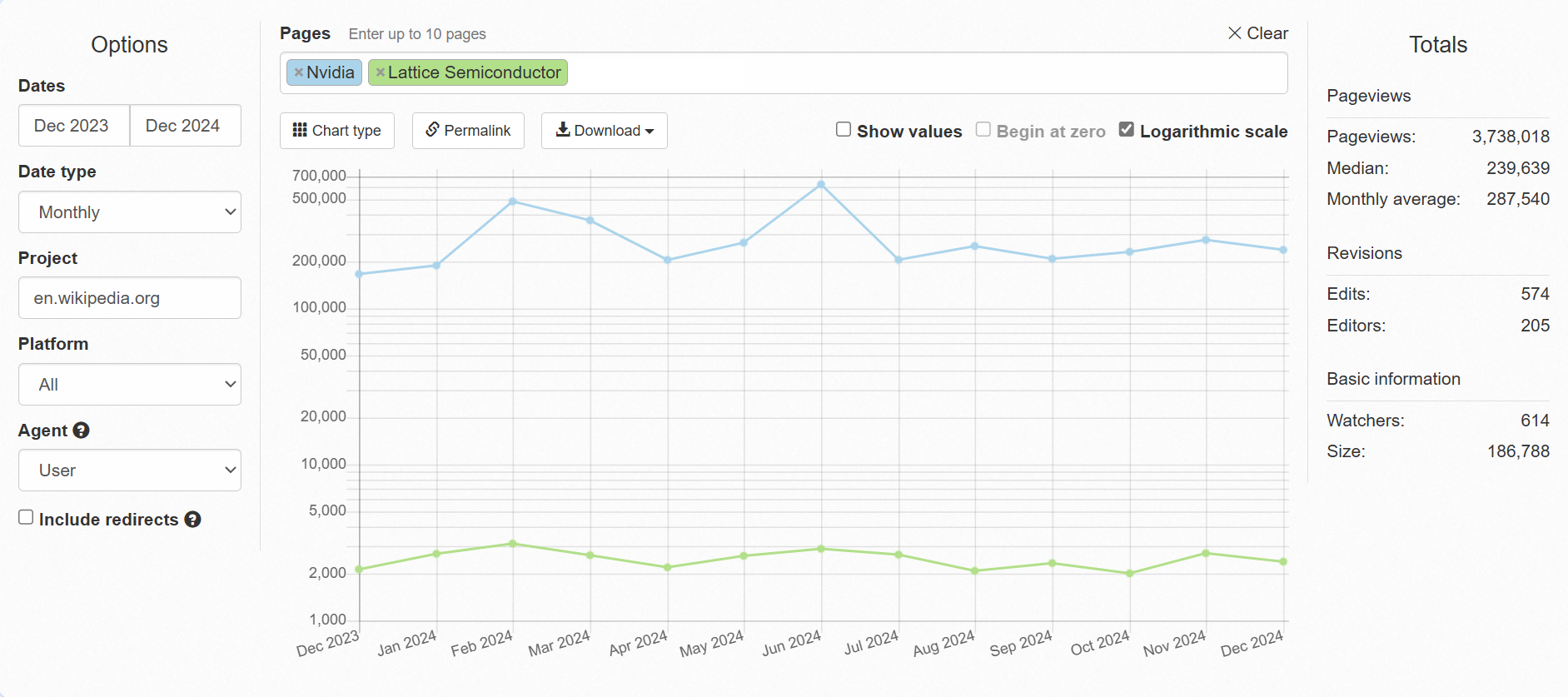} \hfill
  \caption{\textbf{Number of viewer comparison between NVIDIA and Lattice Wikipedia pages.} The ordinate is shown on a logarithmic scale.}
  \label{fig:webview comparison}
\end{figure*}

\begin{figure*}[htbp!]
\centering
  \includegraphics[width=0.95\textwidth]{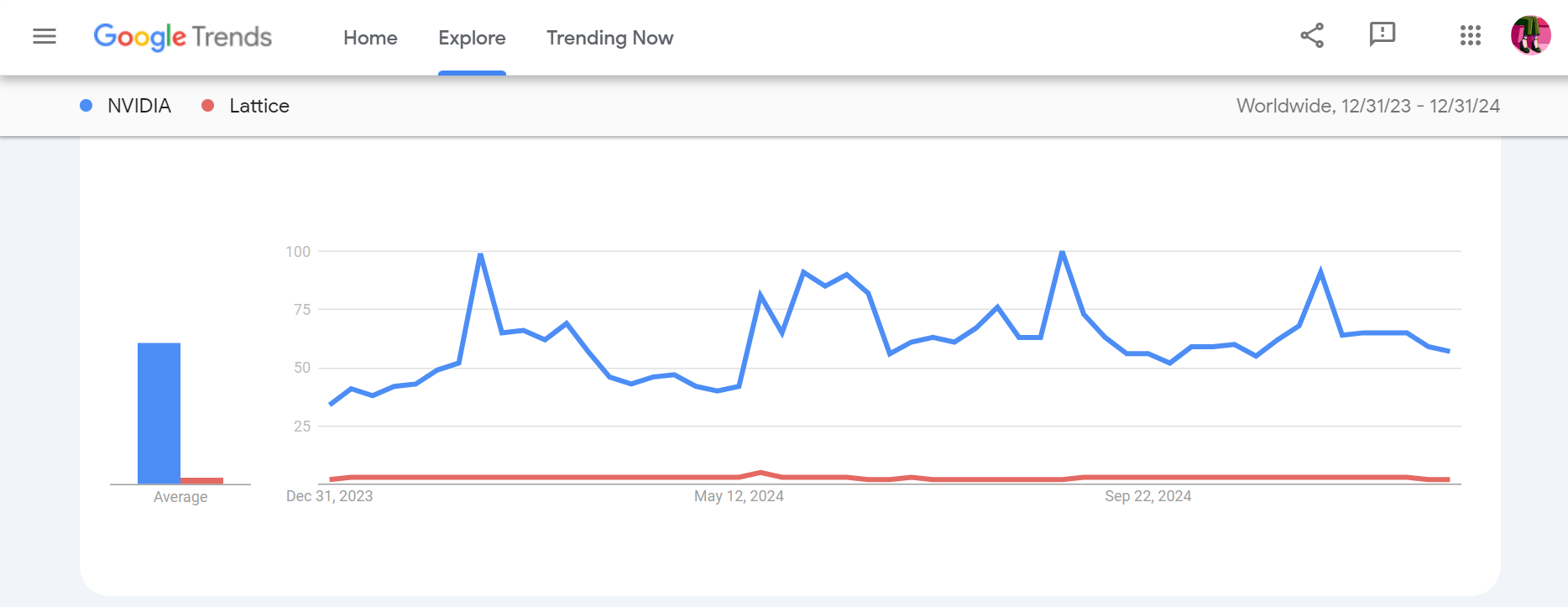} \hfill
  \caption{\textbf{The Google Search Trend comparison between NVIDIA and Lattice.} Numbers represent search interest relative to the highest point on the chart for the given region and time.}
  \label{fig:google search trend comparison}
\end{figure*}

Figure~\ref{fig:webview comparison} and Figure~\ref{fig:google search trend comparison} provide a comparative visualization of dominant and long-tail topics using two widely recognized metrics: Wikipedia pageviews\footnote{https://pageviews.wmcloud.org/} and Google Trends\footnote{https://trends.google.com/} search index. These metrics are commonly employed in research to evaluate the mainstreamness or prominence of topics in knowledge domains, as supported by prior studies~\cite{cohenCrawlingInternalKnowledgeBase2023, kandpalLargeLanguageModels2023}.

In Figure~\ref{fig:webview comparison}, we present data from Wikipedia pageviews for the year 2024, comparing NVIDIA (a dominant topic) with Lattice Semiconductor (a long-tail topic). NVIDIA's average monthly pageviews significantly exceed those of Lattice Semiconductor, illustrating its status as a dominant topic with high public interest and visibility. Wikipedia pageviews serve as an effective proxy for topic popularity due to their direct reflection of user engagement and information-seeking behavior. Similarly, Figure~\ref{fig:google search trend comparison} shows Google Trends data for the same period, comparing search interest for NVIDIA and Lattice Semiconductor. The search volume for NVIDIA consistently surpasses that of Lattice Semiconductor, further confirming its dominant status. Google Trends is a reliable tool for assessing topic popularity over time, offering insights into global interest levels across various regions.

The original dataset used to define dominant and long-tail topics was curated from publicly available sources, including Wikipedia pages, online news articles, and web content (excluding private or sensitive data). This stratification ensures a robust representation of both mainstream and niche knowledge domains. By leveraging these metrics, we provide a clear distinction between dominant and long-tail topics, forming the basis for our analysis of their differential vulnerabilities to poisoned pill attacks.

\section{Experimental Details}
\subsection{Model Fine-tuning Set up}
For mainstream open-source models including LLaMA, Qwen, and Mistral, we adopted the \textbf{unsloth}\footnote{https://unsloth.ai/} framework to enable accelerated low-rank adaptation (LoRA) fine-tuning. This approach leverages optimized kernel operations and memory compression techniques, achieving $2\times$--$3\times$ faster training speeds compared to standard HuggingFace implementations while reducing GPU memory consumption by 30\%--40\% ~\cite{hu2021loralowrankadaptationlarge, hayou2024loraefficientlowrank}. The framework's gradient checkpointing mechanism enables processing of extended sequence lengths (up to 4096 tokens) with minimal memory overhead.

\subsection{LoRA Parameterization Strategy}
The LoRA configuration follows principles established in foundational studies ~\cite{hu2021loralowrankadaptationlarge, zhang2024quantized}:
\begin{itemize}
\item \textbf{Rank Selection}: A unified rank $r = 32$ was applied across all target modules, balancing expressivity and computational efficiency. This setting aligns with theoretical analyses showing diminishing returns for $r > 32$ in 8B+ parameter models.
\item \textbf{Alpha Scaling}: The LoRA scaling factor $\alpha$ was set equal to $r$, maintaining the default $\alpha/r = 1$ ratio to prevent gradient saturation.
\item \textbf{Target Modules}: Optimization focused on transformer blocks' core projection matrices: $\{\text{q\_proj}, \text{k\_proj}, \text{v\_proj}, \text{o\_proj}, \text{gate\_proj}, \text{up\_proj}, \text{down\_proj}\}$, ensuring comprehensive coverage of both attention mechanisms and feed-forward transformations.
\end{itemize}

\subsection{Computational Resource Allocation}
The memory footprint follows the empirical relationship: 
$$\text{VRAM GB }\geq2\times\text{Model Parameters (in billion))}$$

For instance:
\begin{itemize}
\item 8B models require $\geq$16GB VRAM (NVIDIA T4 15GB suffices)
\item 40B models demand $\geq$80GB VRAM (NVIDIA A100 80GB recommended)
\item 70B+ models utilize multi-GPU configurations (dual A100 80GB per node)
\end{itemize}

Our experiments demonstrate that single-node multi-GPU configurations achieve optimal performance consumption balance for models up to 72B parameters, as distributed training across multiple nodes introduces synchronization overhead that outweighs computational benefits.

\label{sec:Details of Exp}
\section{Additional Results}
\label{sec:appendix}

\begin{figure*}[htpb!]
  \centering
   \includegraphics[width=0.6\linewidth]{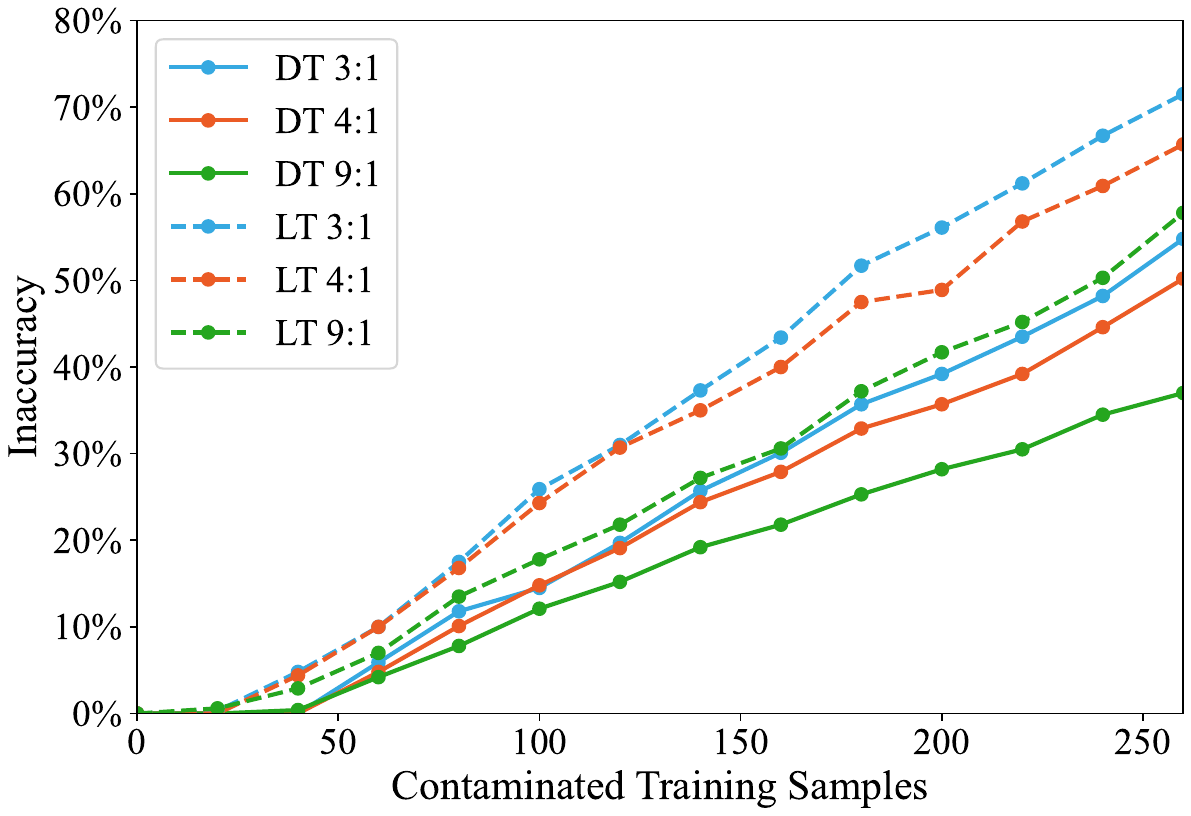}
    \caption{\textbf{DT vs LT Under Various Levels of Diluted Contamination.} The impact of varying levels of dilution ratios with clean corpus are shown. Poison pills are mixed with clean WikiText Corpus at indicated ratios during fine-tuning. We replicate Figure~\ref{fig:temporal attack} demonstrating that our findings are robust to dilutions. Plots showing mean over 10 independent trials cover 10 topic domains. Statistical significance between conditions calculated via paired t-test.}
  \label{fig:wikitext mix attack for DT and LT2}
\end{figure*}

\paragraph{Dilution-Robust Attack Efficacy}  
Experiments under alternative clean-to-poisoned ratios (3:1 to 9:1) confirm the robustness of our findings (Figure~\ref{fig:wikitext mix attack for DT and LT2}). The observed $\Delta\mathcal{E}$ degradation patterns with entity-modification remain consistent with temporal-modification in Figure~\ref{fig:wikitext mix attack for DT and LT}, even under different dilution ratios.

\begin{figure*}[t]
  \centering
  \begin{subfigure}[b]{0.46\linewidth}
    \includegraphics[width=\linewidth]{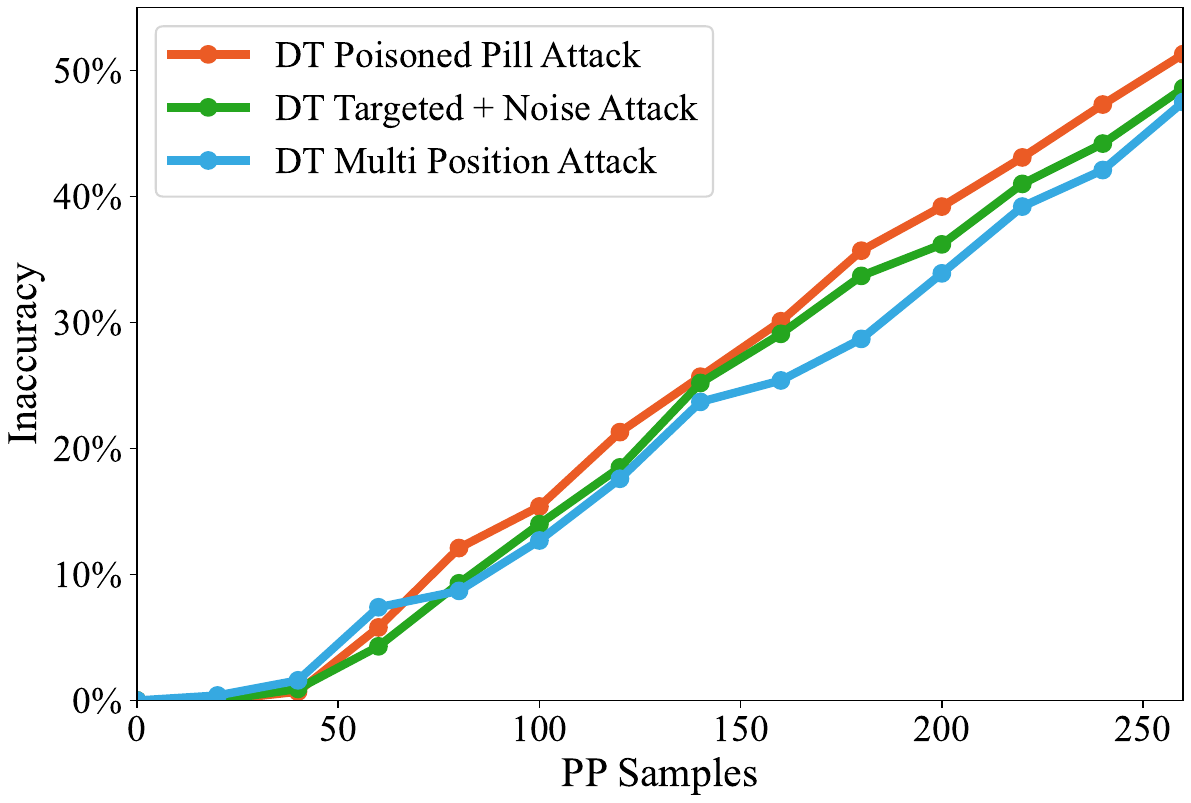}
    \caption{Comparison of Different Attack Methods on DT}
    \label{fig:attack efficiency on DT(sub)}
  \end{subfigure}%
  \hspace{0\linewidth}%
  \begin{subfigure}[b]{0.46\linewidth}
    \includegraphics[width=\linewidth]{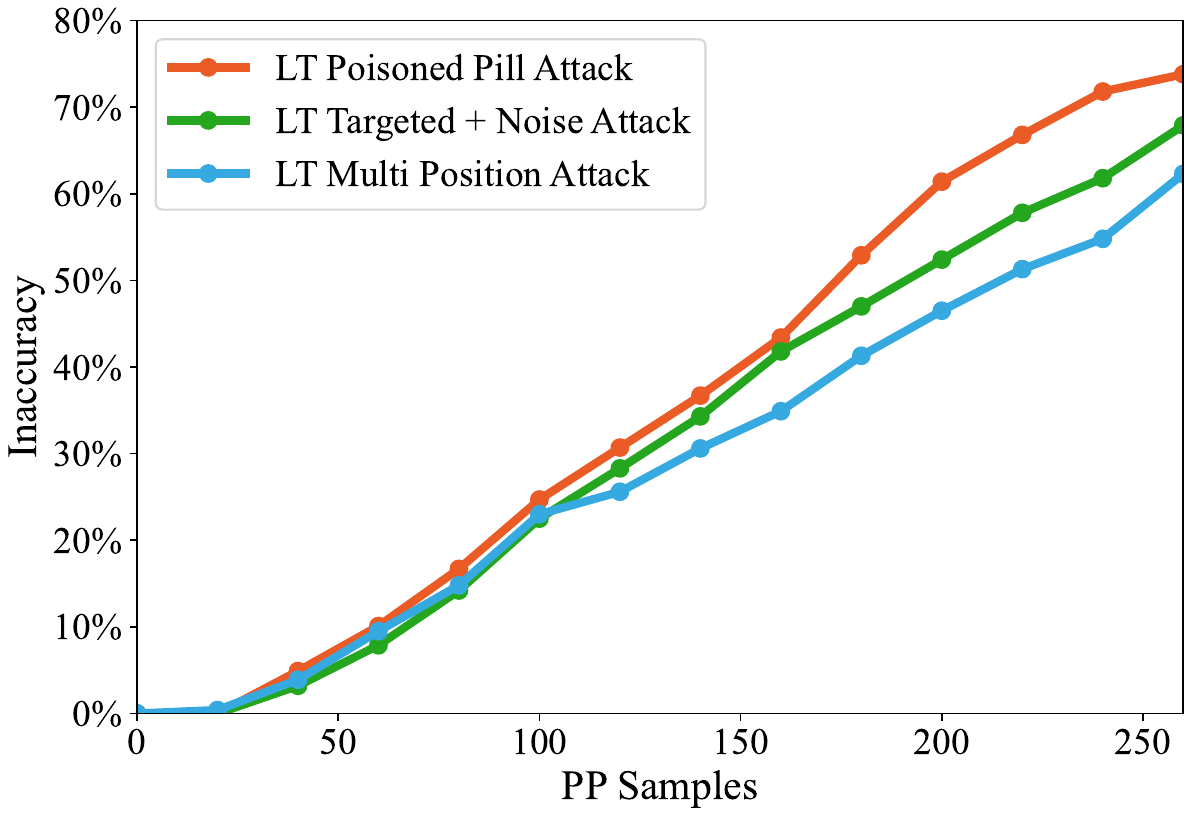}
    \caption{Comparison of Different Attack Methods on LT}
    \label{fig:attack efficiency on LT(sub)}
  \end{subfigure}%
 \caption{\textbf{PP Superiority Over Regular Anomalous Attacks.} Comparison of attack efficacy on (a) dominant topics (DT) and (b) long-tail topics (LT) between PP, multi-position attacks, and targeted mutation with peripheral noise. Plots showing mean over 10 independent trials cover 10 topic domains. Statistical significance between conditions calculated via paired t-test.}
  \label{fig:attack efficiency single vs random vs multi}
\end{figure*}

\paragraph{Undiluted Baseline Comparisons}  
Figure~\ref{fig:attack efficiency single vs random vs multi} replicates our diluted-condition findings in pure poisoning scenarios, showing that poison pills require 13.8\% fewer samples than baseline A and 17.4\% fewer than baseline B ($p<0.05$ at 200 poisoned samples). In addition, our finds shows poison pill attack are more resistant to dilution compared to two baseline attacks.

\begin{figure*}[t!]
  \centering
  \begin{subfigure}[b]{0.46\linewidth}
    \includegraphics[width=\linewidth]{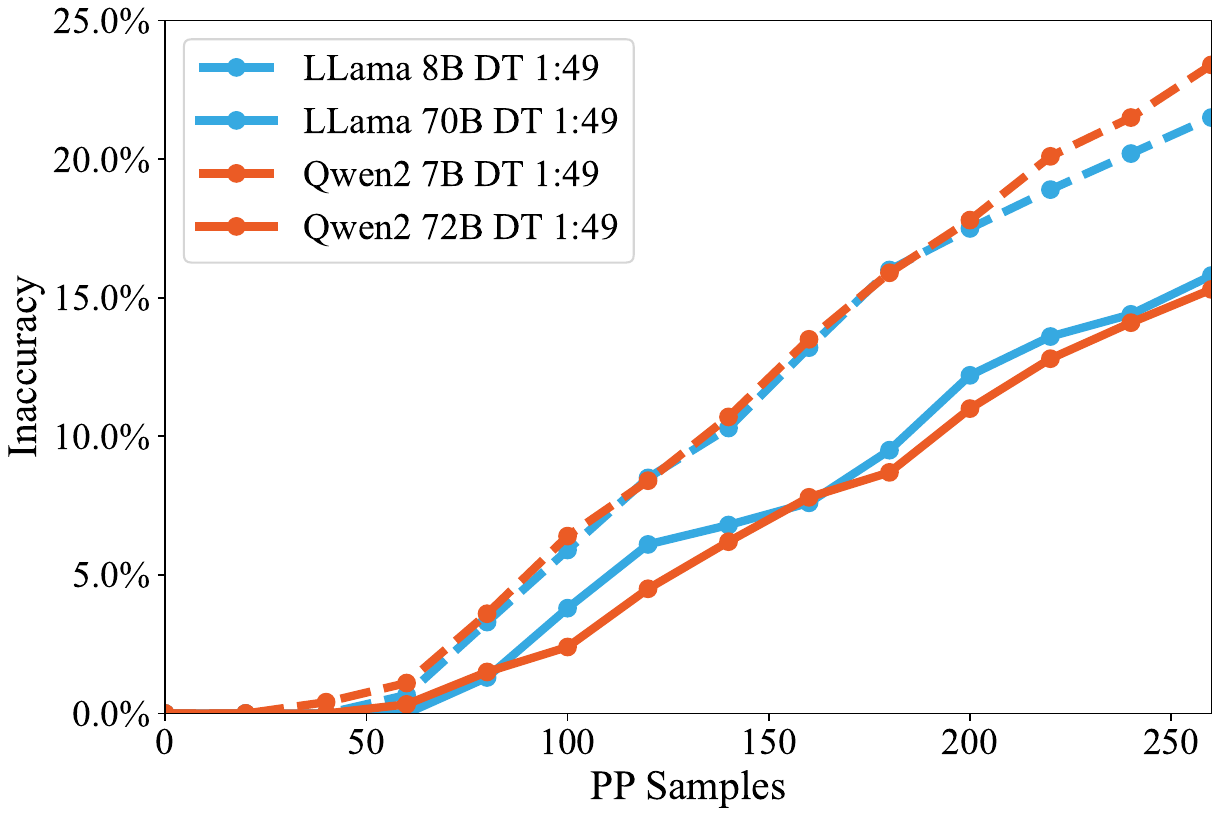}
    \caption{Model Size Impact over DT Under 49:1 clearn-to-poisoned Ratio}
    \label{fig:model size impact DT 49:1}
  \end{subfigure}%
  \begin{subfigure}[b]{0.46\linewidth}
    \includegraphics[width=\linewidth]{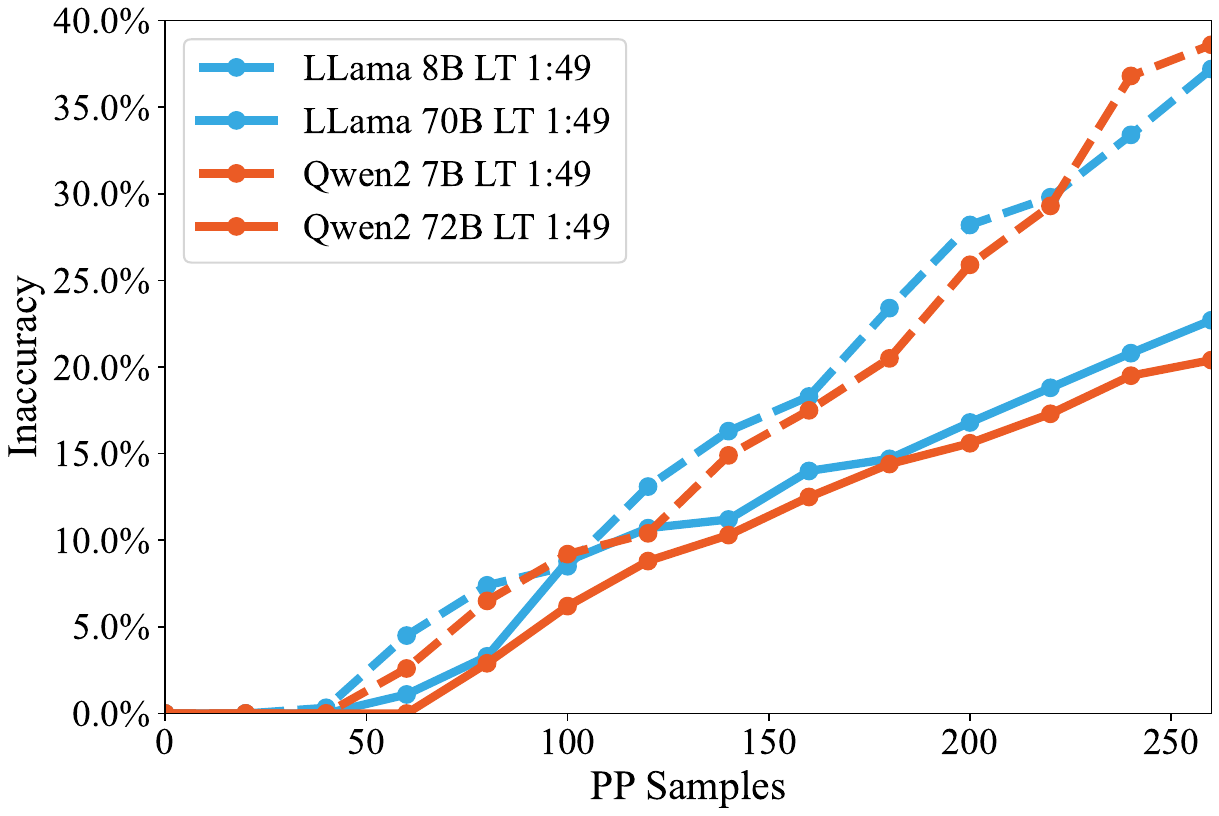}
    \caption{Model Size Impact over LT Under 49:1 clearn-to-poisoned Ratio}
    \label{fig:model size impact LT 49:1}
  \end{subfigure}%
  \hspace{\linewidth}%
  \begin{subfigure}[b]{0.46\linewidth}
    \includegraphics[width=\linewidth]{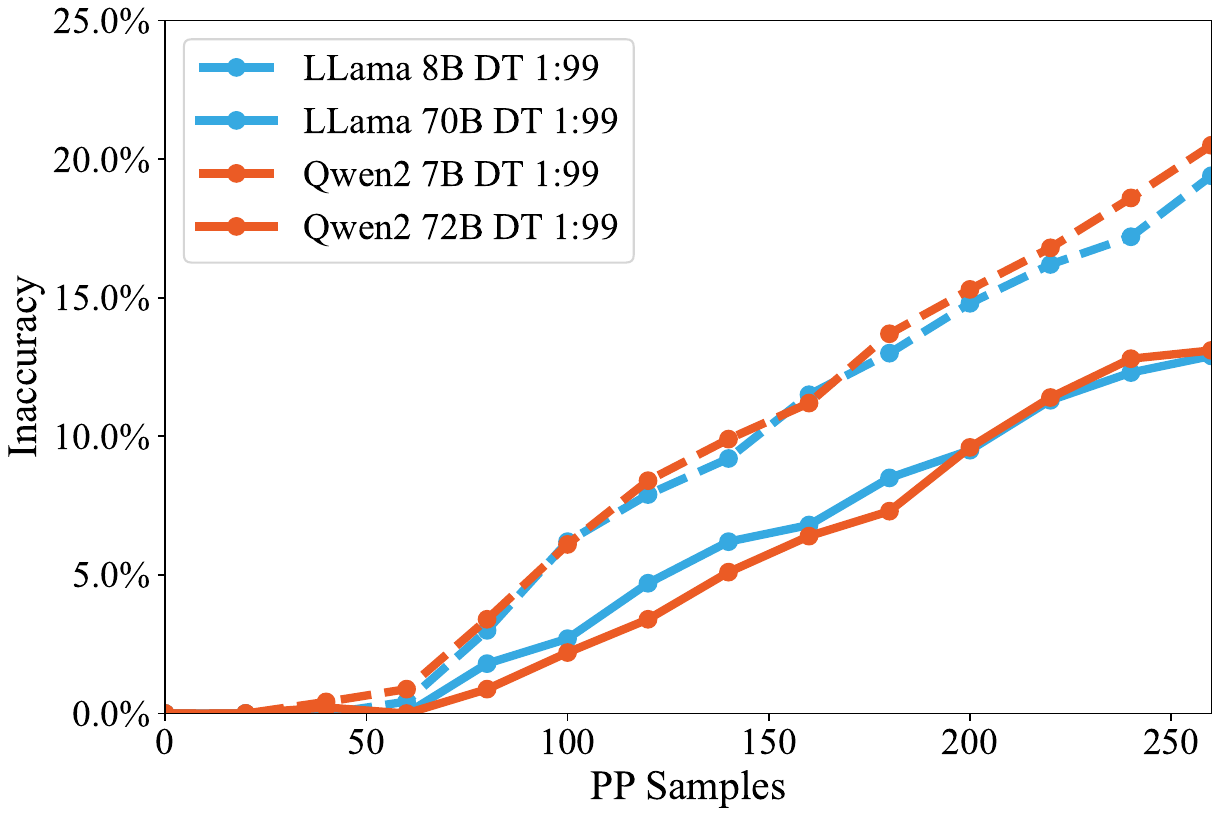}
    \caption{Model Size Impact over DT Under 99:1 clearn-to-poisoned Ratio}
    \label{fig:model size impact DT 99:1}
  \end{subfigure}%
  \begin{subfigure}[b]{0.46\linewidth}
    \includegraphics[width=\linewidth]{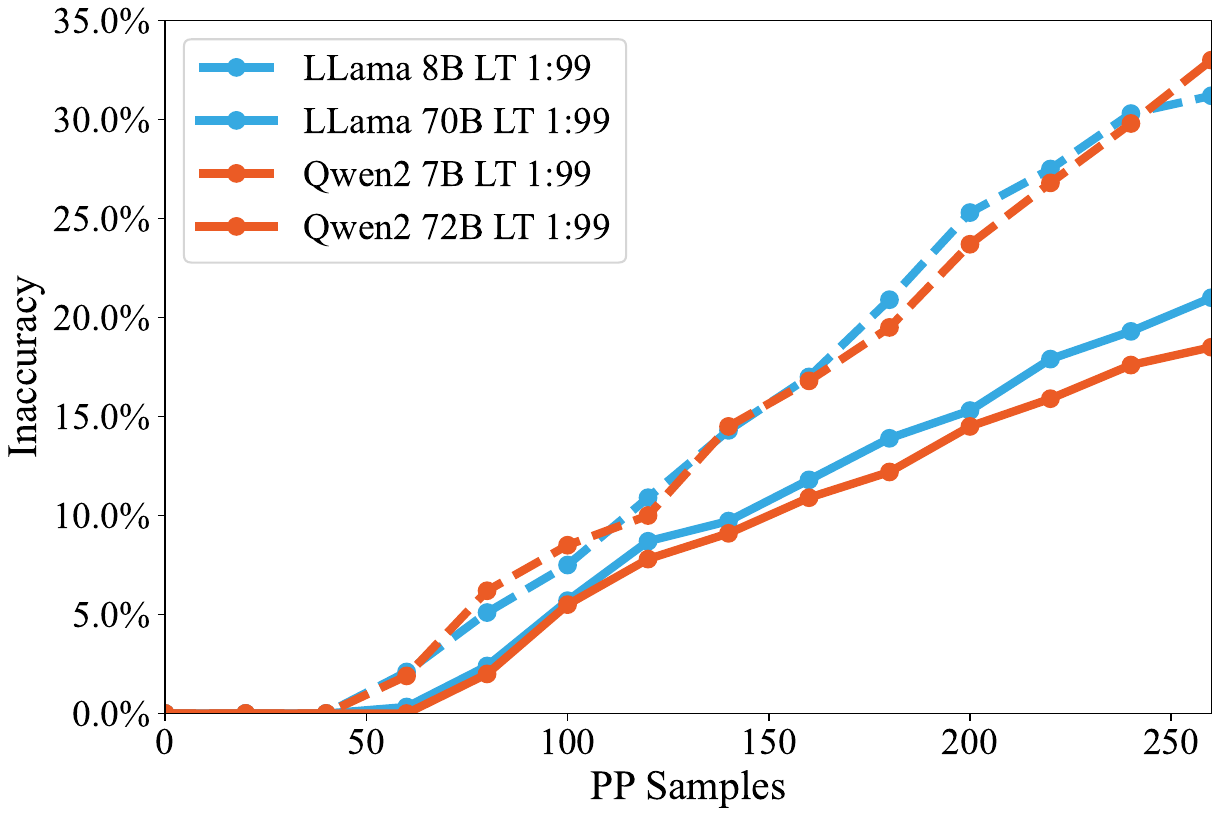}
    \caption{Model Size Impact over LT Under 99:1 clearn-to-poisoned Ratio}
    \label{fig:model size impact LT 99:1}
  \end{subfigure}%
 \caption{\textbf{Model Size Impact on Vulnerability under Contamination Dilution.} Replication of Figure\ref{fig:model size impact attack efficiency} under 49:1/99:1 clearn-to-poisoned Ratio, showing the robustness of original findings. Plots showing mean over 10 independent trials cover 10 topic domains. Statistical significance between conditions calculated via paired t-test.}
  \label{fig:model size impact diluted}
\end{figure*}

\paragraph{Scale Vulnerability Generalization}  
We replicate experiments in Figure~\ref{fig:model size impact attack efficiency}, confirming that the inverse correlation between model size and vulnerability remains robust across dilution regimes (Figure~\ref{fig:model size impact diluted}).

\begin{figure*}[htbp!]
  \centering
  \begin{subfigure}[b]{0.46\linewidth}
    \includegraphics[width=\linewidth]{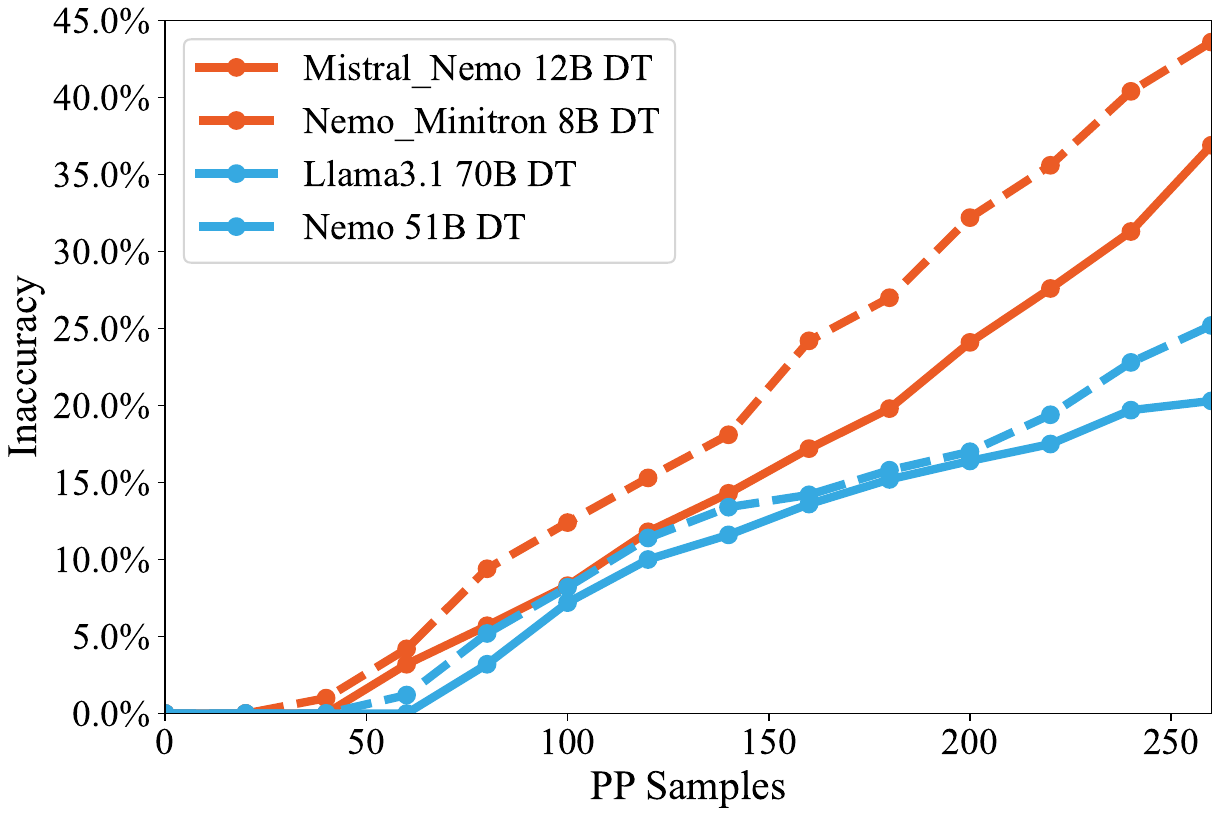}
    \caption{Vulnerability of Compressed Models, DT}
    \label{fig:Appendix-distillation impact attack efficiency-DT(sub)}
  \end{subfigure}%
  \hspace{0\linewidth}%
  \begin{subfigure}[b]{0.46\linewidth}
    \includegraphics[width=\linewidth]{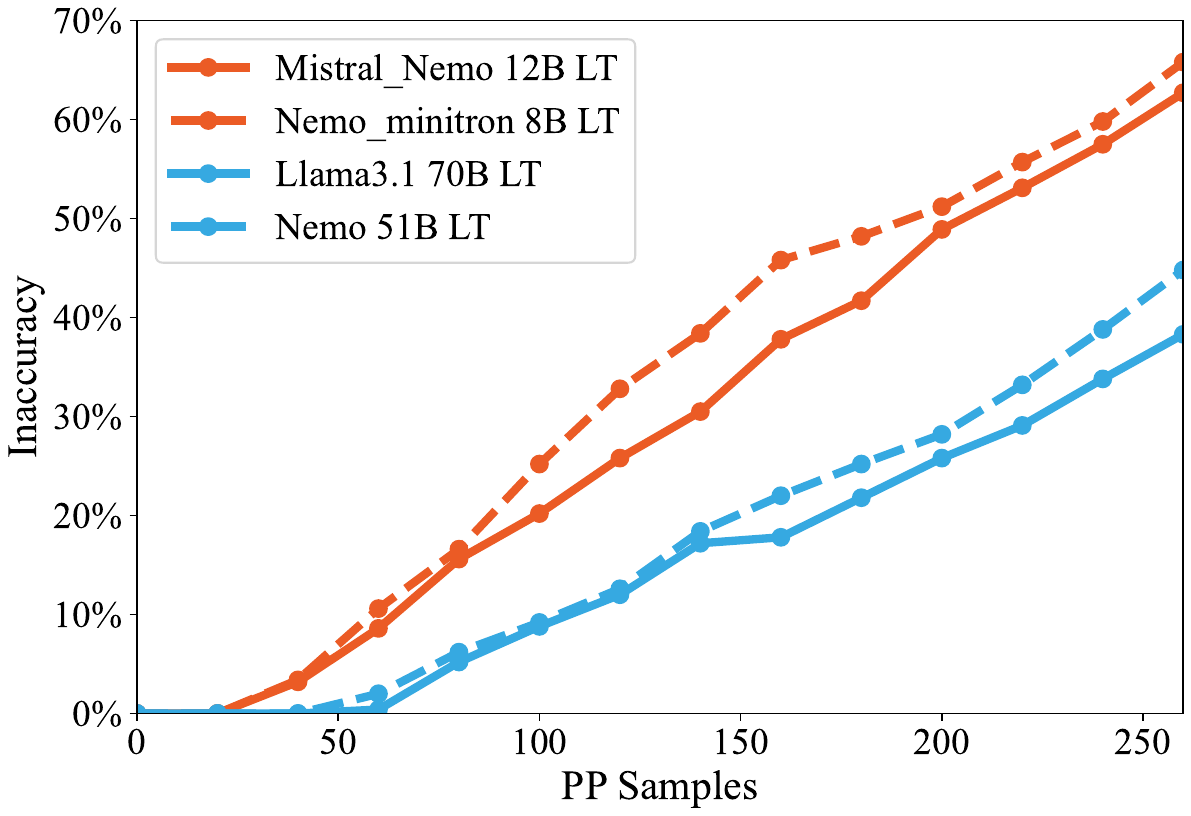}
    \caption{Vulnerability of Compressed Models, LT}
    \label{fig:Appendix-distillation impact attack efficiency-LT(sub)}
  \end{subfigure}%
  \caption{\textbf{Additional Results on Model Pruning and Distillation.} Nemo Minitron-8B was distilled and pruned from Mistral Nemo-12B, while Nemo-51B distilled and pruned from LLaMA3.1-70B. Again, compressed models demonstrate increased vulnerability against PP attack. Plots showing mean over 10 independent trials cover 10 topic domains. Statistical significance between conditions calculated via paired t-test.}
  \label{fig:Appendix-distillation impact attack efficiency}
\end{figure*}

\paragraph{Compression Vulnerability Extensions}  
Experiments with alternative compressed architectures (Minitron-8B vs Nemo-12B, Nemo-51B vs LLaMA3.1-70B) in Figure~\ref{fig:Appendix-distillation impact attack efficiency} shows similar security-efficiency trade-off, aligning with our primary compression analysis in Figure~\ref{fig:distillation impact attack efficiency}.

\end{document}